\documentclass[aps, pra, superscriptaddress, showpacs,
twocolumn, floatfix, 10pt
]{revtex4-1}

\usepackage{amsmath, amssymb}
\usepackage{slashed}
\usepackage{nicefrac}
\usepackage{graphicx}
\usepackage{subfigure}
\usepackage{mathrsfs}

\usepackage{mathptmx}

\usepackage[colorlinks, citecolor=blue, linkcolor=blue, urlcolor=blue]{hyperref}

\newcommand{\figref}[1]{Fig.\,\ref{fig:#1}}
\newcommand{\tabref}[1]{Tab.\,\ref{tab:#1}}
\newcommand{\equref}[1]{Eq.\,\eqref{eq:#1}}
\newcommand{\secref}[1]{Sec.\,\ref{sec:#1}}

\newcommand{\figlab}[1]{\label{fig:#1}}
\newcommand{\tablab}[1]{\label{tab:#1}}
\newcommand{\equlab}[1]{\label{eq:#1}}
\newcommand{\seclab}[1]{\label{sec:#1}}

\renewcommand{\vec}{\mathbf}

\newcommand{\im}{\mathrm{i}}
\newcommand{\e}{\mathrm{e}}
\newcommand{\diff}{\mathrm{d}}
\newcommand{\one}{1}
\newcommand{\degree}{^{\circ}}
\newcommand{\smat}{\mathcal{S}}
\renewcommand{\smat}{\mathscr{S}}

\newcommand{\sci}[2]{{#1}{\times}{10^{#2}}}
\newcommand{\unit}[1]{~\text{#1}}
\newcommand{\abs}[1]{\left|\,{#1}\,\right|}
\newcommand{\vecprod}[2]{{#1}\!\cdot\!{#2}}
\newcommand{\scalar}[3][\mu]{#2_#1 #3^#1}

\newcommand{\avg}{\overline}
\newcommand{\compl}{\bar}
\newcommand{\quot}[1]{``{#1}''}

\begin{document}

\title{Interference Effects in Bethe--Heitler Pair Creation in a Bichromatic Laser Field}

\author{Sven Augustin}
\email{sven.augustin@mpi-hd.mpg.de}

\author{Carsten M\"uller}
\email{carsten.mueller@tp1.uni-duesseldorf.de}

\affiliation{Institut f\"ur Theoretische Physik I, Heinrich-Heine-Universit\"at D\"usseldorf, Universit\"atsstr. 1, 40225 D\"usseldorf, Germany}
\affiliation{Max-Planck-Institut f\"ur Kernphysik, Saupfercheckweg 1, 69117 Heidelberg, Germany}

\date{\today}

\begin{abstract}
We study the creation of electron-positron pairs in the superposition of a nuclear Coulomb field and a two-color laser field of high intensity. Our focus lies on quantum interference effects, which may arise if the two laser frequencies are commensurable.
We show that the interference manifests in the angular distributions of the created particles, which are discussed in the nuclear rest frame and the laboratory frame.
Additionally, we demonstrate that the total pair-production rates can be affected by interference and identify the relative phase between the two laser modes, which optimizes the pair-production yield.
\end{abstract}

%\keywords{}

\maketitle
%\tableofcontents

\section{Introduction}

Creation of electron-positron ($e^- e^+$) pairs by a highly energetic photon in the presence of a nuclear Coulomb field is referred to as the Bethe--Heitler process \cite{bethe-heitler}. Its first observations in a laboratory relied on $\gamma$-rays from nuclear decays or bremsstrahlung \cite{BHexp1, *BHexp2, *BHexp3}. Nowadays, the process serves for applications such as the generation of polarized positron beams \cite{omori, *alexander}, which are of relevance for experimental particle physics.

When a nucleus is subject to the intense photon field of a high-power laser beam, a multiphoton generalization of the usual Bethe--Heitler process may occur, often referred to as the nonlinear Bethe--Heitler process. Here, $n$ laser photons are absorbed simultaneously from the laser field in order to overcome the pair production threshold:
\begin{equation} %\label{bethe-heitler}
Z + n \omega \to Z + e^- + e^+.
\end{equation}

Theoretical investigations of this and similar \emph{matter from laser light} reactions \cite{yakovlev, reiss, nikishov} are almost as old as the demonstration of the first laser itself \cite{maiman}. The interest has been strongly revived in recent years \cite{ehlotzky-review, dipiazza-review}. This development is stimulated by the large and still ongoing progress in high-intensity laser technology. Apart from that, a pioneering experiment at the Stanford Linear Accelerator Center (SLAC) demonstrated that $e^-e^+$ pair production by multiphoton absorption is feasible in collisions of a highly energetic electron beam with an intense laser pulse \cite{e144-1997, *e144-1999}. In a similar manner, the nonlinear Bethe--Heitler reaction is accessible by modern experimental techniques, e.g., by using the highly relativistic nuclear beam from the Large Hadron Collider (LHC) at CERN in conjunction with a counterpropagating high-intensity laser beam. In the nuclear rest frame, the laser frequency and intensity are largely amplified by a relativistic Doppler shift, reaching the levels required for pair production.

The prospect of an experimental test naturally fuels further theoretical investigations. Thus, a substantial amount of work dedicated to the nonlinear Bethe--Heitler effect has been done. Production rates and particle spectra in various field parameter regimes have been calculated (e.g., \cite{cm-circ, avetissian, cm-lin, sieczka, kuchiev, fillion}), and more refined properties were examined, such as nuclear recoil \cite{smueller, krajewska-recoil} and electron-spin effects \cite{tomueller}. It should be noted that in these studies the laser field was always assumed to be a monochromatic plane wave.

In the interaction of a single photon with a structured target Bethe--Heitler pair creation may exhibit signatures of quantum interference. In particular, coherently enhanced pair creation by a photon propagating through a crystal has been studied in detail \cite{palazzi, *uggerhoj}. Similar interference effects occur in pair production on molecules \cite{proriol, voitkiv}. In both cases, the pairs can be produced at two or more Coulombic centers, with the corresponding $\smat$-matrix amplitudes adding up coherently and leading to interference.

In the multiphoton case of the Bethe--Heitler process, a different kind of quantum interference can occur when the laser field is composed of two frequency modes:
\begin{equation} %\label{two-color}
Z + n_1 \omega_1 + n_2 \omega_2\to Z + e^- + e^+.
\end{equation}
If both modes propagate in the same direction and have commensurable frequencies, i.e., frequencies with a rational ratio, it may happen that the total four-momentum of $n_1$ photons from the first mode equals that of $n_2$ photons from the second mode. With this condition fulfilled, it is indistinguishable whether a pair was produced through photon absorption from the first or the second mode. Thus, these two quantum paths can interfere. A similar kind of two-color quantum interference is well known for photoinduced atomic processes \cite{ehlotzky} and chemical reactions \cite{shapiro}, where it can be exploited for coherent control schemes.

Recently, the nonlinear Bethe--Heitler process has been investigated in a bichromatic laser field of commensurable frequencies, with both modes being linearly polarized along the same direction \cite{krajewska-phase, krajewska-symm}. In addition, both modes were assumed to have the same value for the intensity parameter (defined in \equref{xi} below). There, the relative phase between the modes was shown to exhibit a distinct influence on the angular distributions of the created particles. The nonlinear Bethe--Heitler process was also considered for a bichromatic laser field of commensurable frequencies and circularly polarized modes \cite{roshchupkin} and for the case of largely differing, non-commensurable frequencies \cite{dipiazza}. A related study revealed interference effects in $e^-e^+$ pair creation by a highly energetic non-laser photon in the presence of a bichromatic laser field of commensurable frequencies \cite{narozhny}. Finally, it is worth mentioning that other types of interference effects in field-induced pair production have been subject to theoretical investigation recently as well \cite{dumlu-Dunne, *akkermans-Dunne, cheng-grobe, jiang-grobe}.

In the present paper, we study the nonlinear Bethe--Heitler effect in a bichromatic laser field of commensurable frequencies. Both field modes are assumed to be linearly polarized with mutually orthogonal polarization vectors.
%
%This way our study complements earlier investigations where different field geometries were considered \cite{krajewska-phase, krajewska-symm, roshchupkin}.
%
Our theory relies on an $\smat$-matrix formalism in the Furry picture using Volkov solutions to the Dirac equation as basis states. The nuclear Coulomb field is treated in the lowest order of perturbation theory.

Our focus lies on signatures of two-color quantum interference in the pair production process. To this end, the intensity ratio of the frequency modes will be chosen in a way to maximize the interfering contributions in the square of the $\smat$-matrix. We will show that the two-color interference modifies the angular distribution of the created particles. In particular, shifts of the angular peak positions are found. Under certain conditions, the interference may also lead to an increase (or a decrease) of the total pair-production rate. These changes are shown to depend on the relative phase between the two field modes. Thus, the latter can be chosen to maximize the yield of produced pairs. The results are discussed in the nuclear rest frame and the laboratory frame. Finally, an intuitive explanation for the phase dependence of the total pair production rate is developed.

Regarding the chosen geometry of the bichromatic laser field, we note that the orthogonality of the field modes offers two advantages. On the one hand, it simplifies the mathematical treatment of the process due to the vanishing of certain cross terms. For the same reason it guarantees that, on the other hand, the laser intensity remains unchanged under variation of the relative phase between the modes. Both features facilitate gaining intuitive insights, such as the one mentioned above, into the rather complex nature of two-color quantum interferences in the nonlinear Bethe--Heitler effect. Furthermore, in this context it is interesting to note that orthogonally polarized two-color laser fields have proven to be beneficial in atomic physics in order to control the laser-driven recollision dynamics of field-ionized electrons \cite{kitzler, brugnera}.

This paper is organized as follows. First, we will outline our calculational approach in \secref{calc}. The Volkov solutions of the Dirac equation in a bichromatic laser field with linearly polarized modes of orthogonal field vectors will be given. Afterwards, the $\smat$-matrix describing the nonlinear Bethe--Heitler process in such a laser field is evaluated, and an expression for the total pair-production rate is derived. The latter contains a six-fold integral over the momenta of the created particles and a four-fold sum over photon numbers, which both can effectively be reduced by one due to energy constraints. The remaining integrations are performed numerically, leading to the results presented in \secref{results}, where we show pair production rates differential in the polar emission angle for various frequency ratios and (total) photon energies. The conclusions that can be drawn from our study are summarized in \secref{concl}.

\section{Theoretical Framework} \seclab{calc}
\subsection{Volkov Solutions and Field Geometry}

For an electron moving in the field of an electromagnetic plane wave in vacuum the Dirac equation can be solved exactly, leading to the so-called Volkov solutions, which were first derived in 1935 \cite{volkov}. A brief overview of this derivation, with special emphasis on the case of a bichromatic laser wave, is given in the following.

The Dirac equation \cite{dirac}
\begin{equation} \equlab{dirac}
\left( \im \hbar \slashed{\partial} + \frac{e}{c} \slashed{A} - mc \right) \Psi = 0,
\end{equation}
with the positive elementary charge $e$, applying Feynman slash notation $\slashed{A} = \scalar{\gamma}{A}$ and defining the four-gradient $\partial = \left( \frac{1}{c}\frac{\partial}{\partial t}, -\nabla \right)$, can be solved analytically for a vector potential $A(\eta)$ depending only on a phase variable
\begin{equation}
\eta = \scalar{k}{x} = \omega t - \vecprod{\vec{k}}{\vec{r}},
\end{equation}
and given in Lorenz gauge $\scalar{\partial}{A} = 0$, which corresponds to $A$ being transversal: $\scalar{k}{A} = 0$.
Here the wave vector $k = (\nicefrac{\omega}{c}, \vec{k})$ and the four-dimensional space-time coordinate $x = (ct, \vec{r})$ are used.
%Scalar products of two four-vectors are denoted by implied summation over repeated indices.
%
This leads to the Volkov solutions for electrons and positrons, as denoted by the superscripts $(-)$ and $(+)$, respectively:
\begin{equation} \equlab{volkov}
\Psi_{p,s}^{(\pm)} = N 
\left( \one \pm \frac{e \slashed{k} \slashed{A}}{2c \scalar{k}{p}} \right)
\exp\!{\left( \frac{\im}{\hbar} S^{(\pm)} \right)} ~u_{p,s}^{(\pm)}.
\end{equation}
Here $N$ is a normalizer, $u$ a free Dirac spinor with the respective particle's momentum $p$ and spin $s$, and $S$ the action as given by
\begin{equation} \equlab{action}
S^{(\pm)} = \pm \scalar{p}{x} + \frac{e}{c\scalar{p}{k}} \int^\eta \! \left[ \scalar{p}{A}(\tilde\eta) \mp \frac{e}{2c} A^2(\tilde\eta) \right] \diff\tilde\eta.
\end{equation}
Using these wave functions, the interaction between electrons or positrons and the laser field can be treated exactly.

In the present study, the laser field $A$ is defined as superposition of two laser waves: $A = A_1 + A_2$. For each of the both laser modes, we assume a general plane wave field:
\begin{equation} \equlab{laserwave}
A_i = a_i \cos(\eta_i + \varphi_i)
\quad
(i=1 ~\text{or}~ 2),
\end{equation}
with perpendicular field vectors $a_i$, relative phases $\varphi_i$, and phase coordinates $\eta_i = (\nicefrac{\omega_i}{c}) \, \scalar{\kappa}{x}$, where both $k_i = (\nicefrac{\omega_i}{c}) \kappa$ share the same direction of propagation $\kappa = \left( 1, 0, 0, 1 \right)$, as depicted in \figref{wave}.
%
%Without loss of generality, we can simplify our treatment by keeping only one relative phase $\varphi = \varphi_1$ and setting $\varphi_2 = 0$ for all cases.
%
The field vectors $a_i$ are given by
\begin{equation}
a_1 = \left( 0, 1, 0, 0 \right) \abs{\vec{a}_1\!}
\quad\text{and}\quad
a_2 = \left( 0, 0, 1, 0 \right) \abs{\vec{a}_2\!}.
\end{equation}
The dimensionless intensity parameters
\begin{equation} \equlab{xi}
\xi_i = \frac{e}{mc^2} \frac{\abs{\vec{a}_i\!}}{\sqrt{2}}
\end{equation}
can be used to measure their absolute amplitudes.

\begin{figure}[t]
\begin{center}
\includegraphics[width=\columnwidth]{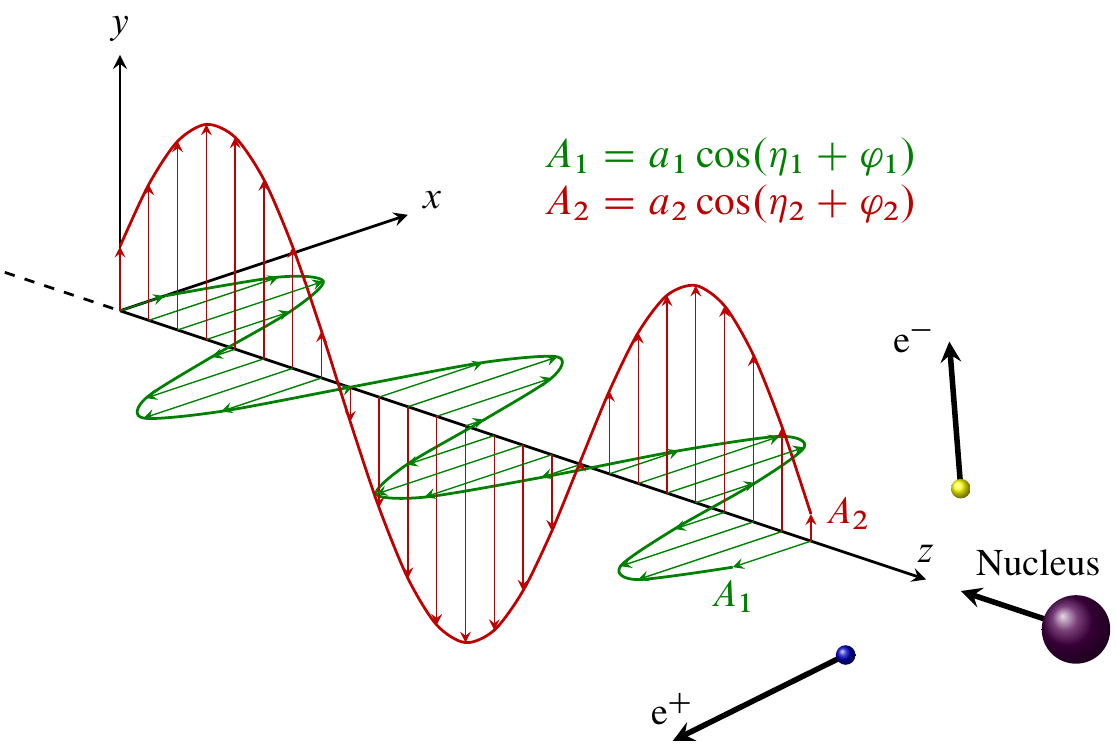}
\caption{\figlab{wave}(Color online)~
The applied pair creation scheme --
Two linearly polarized plane waves with perpendicular field vectors shined in along the $z$-axis onto a counter-propagating nucleus, creating an electron-positron pair.
}
\end{center}
\end{figure}

In the superposition of these two fields, we can make use of the fact that despite the squared laser amplitude in \equref{action}, functions of the two phase coordinates are separable:
\begin{align} \equlab{actionsum}
S^{(\pm)}
&= \pm \scalar{p}{x} + \sum_{i=1}^{2} S_i^{(\pm)}, \\
\equlab{actionsep}
S_i^{(\pm)}
&=
\frac{e}{c\scalar{p}{k_i}}
\int^{\eta_i}
\!\!
\left[ \scalar{p}{A_i}(\tilde\eta_i) \mp \frac{e}{2c} A_i^2(\tilde\eta_i) \right]
\diff\tilde\eta_i.
\end{align}
It follows that this separability, which represents a special feature of the considered field geometry, also carries over to \equref{volkov}.
Upon insertion of \equref{laserwave} the integral in \equref{actionsep} can be calculated, yielding
\begin{equation} \equlab{actionint}
S_i^{(\pm)} =
\frac{e}{c\scalar{p}{k_i}}
\left[
\scalar{p}{a_i} \sin(\eta'_i)
\mp \frac{e}{4c} a_i^2 \left(
\frac{\sin(2\eta'_i)}{2}
+ \eta'_i
\right)
\right],
\end{equation}
where we have abbreviated $\eta'_i = \eta_i  + \varphi_i$. At this point it is useful to define the effective momentum
\begin{equation}
q = p + \frac{e^2 \avg{\vec{A}^2}}{2c^2 \scalar{p}{\kappa}} \, \kappa,
\end{equation}
with the averaged squared laser amplitude
\begin{equation}
\avg{\vec{A}^2} 
= \frac{1}{2} \left(\abs{\vec{a}_1\!}^2 + \abs{\vec{a}_2}^2\right)
= \frac{m^2 c^4}{e^2} \left(\xi_1^2 + \xi_2^2\right).
\end{equation}
%
%and $\kappa$ being the direction of propagation shared by the two laser waves.
The effective momentum characterizes the electronic motion in the laser field and contains the linear term in $\eta_i$ from \equref{actionint}, while the linear term in $\varphi_i$ will eventually cancel in the following steps and is thus not further discussed. Additionally, we can conclude that due to the orthogonal field vectors the mean intensity of the bichromatic laser field is independent of the relative phases $\varphi_i$.

The definition of the effective momentum also allows us to give an expression for the normalizer in \equref{volkov}:
\begin{equation} \equlab{normalizer}
N = \sqrt{\frac{m c}{q^0}}.
\end{equation}
In addition, the Lorentz-invariant square of the effective momentum
\begin{equation}
q^2 = p^2 + \frac{e^2}{c^2} \avg{\vec{A}^2} 
\end{equation}
may be used to define the effective electron mass
\begin{equation}
m_* = m \sqrt{1 + \xi_1^2 + \xi_2^2}.
\end{equation}

\subsection{Transition Amplitude}

We model pair production as a transition from a negative continuum Volkov state $\Psi^{(+)}$ to one of the positive continuum $\Psi^{(-)}$, induced by a nuclear Coulomb potential with a vanishing vectorial part $\vec{A}_\text{N} = \vec{0}$ and
a scalar part $A_\text{N}^0 = \nicefrac{Ze}{\abs{\vec{r}}}$.
%
%Note that for the results shown in \secref{results} we use $Z = 1$, assuming a proton beam target.
%
We write the nuclear four-potential accordingly as
\begin{equation}
A_\text{N} = \frac{Ze}{\abs{\vec{r}}} \epsilon,
\end{equation}
using $\epsilon = (1,0,0,0)$, leading to the pair-creation amplitude
\begin{equation}
\smat = \frac{\im \, e}{\hbar c} \int \compl\Psi_{p_-,s_-}^{(-)} \slashed{A}_\text{N} \Psi_{p_+,s_+}^{(+)} \, \diff^4 \! x.
\end{equation}
From here on we shall distinguish between electron and positron by the subscripts $-$ and $+$, respectively, for the (effective) momentum, the spin, and the normalizer.
Upon insertion of the Volkov wave functions from \equref{volkov} we obtain
\begin{equation}
\smat = 
N_- N_+ 
\frac{\im Z e^2}{\hbar c} 
\int \! \frac{\diff^4 \! x}{\abs{\vec{r}}} \,
G
\exp\!{\left(\frac{\im}{\hbar} \left(-S^{(-)} + S^{(+)} \right)\right)},
\end{equation}
where we have introduced the abbreviation
\begin{equation} \equlab{slashes}
G = 
\compl{u}_{p_-,s_-}^{(-)}
\left(
\one - \frac{e \slashed{A} \slashed{\kappa}}{2c \scalar{\kappa}{p_-}}
\right)
\slashed{\epsilon}
\left(
\one + \frac{e \slashed{\kappa} \slashed{A}}{2c \scalar{\kappa}{p_{\vphantom{-}\smash{+}}}} \right)
u_{p_+ ,s_+}^{(+)},
\end{equation}
containing all $\gamma$-matrices.
Here we insert the action from \equref{actionsum} and the laser fields from \equref{laserwave}.
With the definitions 
\begin{align}
\alpha_i &=
\frac{e}{\hbar \omega_i}
\left(
\frac{a_{i\mu} p_-^\mu}{\scalar{\kappa}{p_-}}
-
\frac{a_{i\mu} p_+^\mu}{\scalar{\kappa}{p_+}} 
\right), \\
\beta_i &=
\frac{e^2 a_i^2}{8c \hbar \omega_i}
\left(
\frac{1}{\scalar{\kappa}{p_-}}
+
\frac{1}{\scalar{\kappa}{p_+}}
\right),
\end{align}
we find a set of three periodic functions $f_j(\eta_i)$ for each laser mode:
\begin{equation} \equlab{periodic}
\begin{split}
f_1(\eta_i) &= \e^{- \im \alpha_i \sin(\eta_i) - \im \beta_i \sin(2\eta_i)}, \\
f_2(\eta_i) &= f_1(\eta_i) \cos(\eta_i), \\
f_3(\eta_i) &= f_1(\eta_i) \cos^2(\eta_i)
\quad
(i=1 ~\text{or}~ 2).
\end{split}
\end{equation}
These can be expanded into Fourier series according to
\begin{equation} \equlab{fourier}
f_j(\eta_i) = \sum_{n_i} C_{n_i}^{(j)} \e^{-\im n_i \eta_i}
\quad
(j \in \{1,2,3\}),
\end{equation}
where $C_{n_i}^{(j)}$ represents the six coefficients, herein built from generalized Bessel functions $\tilde{J}_{n} (\alpha, \beta)$ \cite{reiss-genbes}, as defined via the ordinary Bessel function $J_{n}(x)$ by
\begin{equation}
\tilde{J}_{n} (\alpha, \beta)=\sum_{l=-\infty}^{\infty} 
J_{n-2l} (\alpha) \, J_{l} (\beta).
\end{equation}
This yields the respective Fourier coefficient for each function in \equref{periodic}:
\begin{equation} \equlab{coefficients}
\begin{split}
C_{n_i}^{(1)} &= \tilde{J}_{n_i} {(\alpha_i, \beta_i)}, \\
C_{n_i}^{(2)} &= \frac{1}{2} \left[
\tilde{J}_{n_i-1} {(\alpha_i, \beta_i)} + 
\tilde{J}_{n_i+1} {(\alpha_i, \beta_i)}
\right], \\
C_{n_i}^{(3)} &= \frac{1}{4} \left[
\tilde{J}_{n_i-2} {(\alpha_i, \beta_i)} + 
2 \tilde{J}_{n_i} {(\alpha_i, \beta_i)} + 
\tilde{J}_{n_i+2} {(\alpha_i, \beta_i)}
\right].
\end{split}
\end{equation}
Thus we can write the amplitude as a summation over two indices, which can be interpreted as counts for the number of photons taken from each of the two modes:
\begin{equation} \equlab{smatrix}
\smat = 
\frac{\im Z e^2 m c}{\hbar c \sqrt{q_-^0 q_+^0}} 
\sum_{n_1, n_2}
M_{p_- p_+}^{(n_1, n_2)}
\int \! \frac{\diff^4 \! x}{\abs{\vec{r}}} \,
\exp\!{\left(\frac{\im}{\hbar} \scalar{x}{Q_{(n_1, n_2)}}\right)}.
\end{equation}
Here we have inserted the normalizers from \equref{normalizer}, as well as introduced the momentum transfer to the nucleus
\begin{equation}
Q_{(n_1, n_2)} = q_+ + q_- - n_1 \hbar k_1 - n_2 \hbar k_2
\end{equation}
and the matrix element $M_{p_- p_+}^{(n_1, n_2)}$, which consists of all slashed quantities and the six Fourier coefficients from \equref{coefficients}.

The four-dimensional integral in \equref{smatrix} can be solved by using the Fourier transform of the Coulomb potential and a representation of the $\delta$-function for the integral in space and time ($x_0 = ct$), respectively \cite{bjorken-drell}:
\begin{align}
&\int \! \diff^3 r \, \frac{1}{\abs{\vec{r}}} \, \e^{-\frac{\im}{\hbar} \vecprod{\vec{Q}\,}{\,\vec{r}}} = \frac{4 \pi \hbar^2}{\abs{\vec{Q}}^2}, \\
&\int \! \diff x_0 \, \e^{\frac{\im}{\hbar} Q_0 \, x_0} = 2 \pi \hbar \, \delta(Q_0).
\end{align}
Note that by definition of $Q_{(n_1, n_2)}^0$ this $\delta$-function ensures energy conservation.

Squaring the amplitude leads to a sum over four indices:
\begin{equation} \equlab{amplitude}
\abs{\smat}^2
=
\sum_{\substack{n'_1, n'_2\\n_1, n_2}}
\mathscr{P}_{(n_1, n_2, n'_1, n'_2)},
\end{equation}
with the addends being the thereby defined partial contributions
\begin{align} \equlab{partcontrib}
\mathscr{P}_{(n_1, n_2, n'_1, n'_2)}
&=
C
\compl{M}_{p_- p_+}^{(n_1, n_2)} M_{p_- p_+}^{(n'_1, n'_2)}
\frac{cT}{Q_{(n_1 n_2)}^4} \, \delta\!\left(Q_{(n_1, n_2)}^0\right),
\nonumber
\\
C
&=
\frac{Z^2 e^4 m^2}{\hbar^2 q_+^0 q_-^0}
~
32\pi^3 \hbar^5.
\end{align}
Here we have used
\begin{align}
n_1 k_1 + n_2 k_2 &= n'_1 k_1 + n'_2 k_2, \\
Q_{(n_1, n_2)} &= Q_{(n'_1, n'_2)},
\end{align}
as enforced by the $\delta$-function,
and introduced the time $T$ from the squared $\delta$-function \cite{bjorken-drell}:
\begin{equation} \equlab{deltasquared}
\left[ 2\pi\hbar \, \delta\!\left(Q_{(n_1, n_2)}^0\right) \right]^2 
= 2\pi\hbar \, \delta\!\left(Q_{(n_1, n_2)}^0\right) cT.
\end{equation}

The product of the two matrix elements $\compl{M}_{p_- p_+}^{(n_1, n_2)}$ and $M_{p_- p_+}^{(n'_1, n'_2)}$ is a rather cumbersome summation of products of Dirac $\gamma$-matrices and thus shall not be shown here. The general structure should be discussed nevertheless: The result can be divided into three parts. Two parts depend on the parameters of the individual laser modes separately and can each be identified with the result found for a single monochromatic laser wave \cite{cm-lin}. In contrast, the third part depends on mixtures of the parameters of the two laser modes and thus consists of the additional terms occurring due to their superposition.

Finally, the partial contributions $\mathscr{P}$ enter the differential partial rates
\begin{equation} \equlab{partrate}
\diff^6 R_{(n_1, n_2, n'_1, n'_2)} = 
\frac{1}{T} \sum_{s_+, s_-}
\mathscr{P}_{(n_1, n_2, n'_1, n'_2)}
\frac{\diff^3 q_-}{(2\pi\hbar)^3} \frac{\diff^3 q_+}{(2\pi\hbar)^3},
\end{equation}
where we also summed over the final spin states and divided by the time $T$ introduced in \equref{deltasquared}.
Using the $\delta$-function in \equref{partcontrib}, we can perform one integration analytically. The remaining integrals are calculated numerically to obtain angular differential and fully integrated partial rates. Additionally, the summation over photon numbers from \equref{amplitude} is performed to find differential and total rates.
The results of these computations are presented in the following section.

It should be noted that, in general, the partial rates from \equref{partrate} are no experimental observables. Particularly, they are not necessarily positive quantities.
Summed-up rates, on the other hand, are always positive and measurable.
Negative partial rates in the four-index sum will lead to a decreased summed-up rate. They may arise from certain index combinations that will subsequently be interpreted as destructive interference.

Concluding this section, we point out that for an appropriate set of parameters, in particular by defining
\begin{equation}
\begin{split}
a &= \abs{\vec{a}_1\!} = \abs{\vec{a}_2}, \\
\omega &= \omega_1 = \omega_2, \\
%\eta &= \eta_1 = \eta_2 = (\nicefrac{\omega}{c}) \, \scalar{\kappa}{x}, \\
\varphi_1 &= \nicefrac{\pi}{2}, ~ \varphi_2 = 0,
\end{split}
\end{equation}
we are also able to reproduce previous calculations for a circularly polarized laser wave \cite{cm-circ}.
%%
%\begin{equation}
%A = a
%%\begin{pmatrix}
%\left(\!
%\begin{smallmatrix}
%0 \\
%\cos{\eta} \\
%\sin{\eta} \\
%0
%\end{smallmatrix}
%\!\right).
%%\end{pmatrix}.
%\end{equation}
%%

\section{Results} \seclab{results}
\subsection{Notation and General Remarks}

In order to help read the following results a short remark on the applied notation should be made.
We denote a pair of laser waves by $(\tilde{n}_1, \tilde{n}_2)$ when it is indistinguishable whether $\tilde{n}_1$ photons were absorbed from the first mode or $\tilde{n}_2$ from the second mode.
Here $\tilde{n}_i$ is the minimal number of photons needed from mode $i$ to create a pair, using no photon from the other mode.
We call $\tilde{n}_1 \hbar \omega_1 = \tilde{n}_2 \hbar \omega_2$ the total photon energy, where the frequencies $\omega_1$ and $\omega_2$ are always assumed to be commensurable.

\begin{table*}[ht]
\caption{\tablab{nota}
Types, conditions, and examples of the terms in the summation from \equref{amplitude}.}
\begin{ruledtabular}
\begin{tabular}{@{\hskip 0.1in}lcr@{\hskip 0.1in}}
Type of term & Condition & Example: $\left(2,4\right)$\\ \hline
Direct & $n_1 = n'_1, n_2 = n'_2 \enskip\text{and only one is not zero}$ & $\left[2,2,0,0\right]$ or $\left[0,0,4,4\right]$\\
Symmetrically mixed & $n_1 = n'_1 \neq 0 \enskip\text{and}\enskip n_2 = n'_2 \neq 0$ & $\left[1,1,2,2\right]$ \\
Interference (Asymmetrically mixed) & $n_1 \neq n'_1 \enskip\text{and}\enskip n_2 \neq n'_2$ & $\left[0,2,4,0\right]$ or $\left[2,0,0,4\right]$\\
\end{tabular}
\end{ruledtabular}
\end{table*}

Likewise, we denote combinations of indices in the sum from \equref{amplitude} by $\left[n_1, n'_1, n_2, n'_2\right]$ and distinguish \emph{direct}, \emph{symmetrically mixed}, and \emph{asymmetrically mixed} terms, where the latter stem from interference (see \tabref{nota} for the exact conditions and some examples).
\emph{Direct} terms are those originating solely from one of the two laser modes and thus would also be visible if the other mode was turned off. They can serve as contribution strength references later on, as they could also be used in an experimental comparison by recording the two laser sources one at a time.
The \emph{symmetrically mixed} terms can be understood as taking a certain number of photons from the first mode and another number from the second. In these cases no interference is involved.
For the \emph{interference} terms, on the other hand, it is, by definition, not obvious how to interpret their index combination as actual photon numbers from the two modes.

The interference terms are sensitive to the relative phases $\varphi_i$ from \equref{laserwave}.
Without loss of generality, we can simplify our treatment by keeping only one relative phase $\varphi = \varphi_1$ and setting $\varphi_2 = 0$ for all cases.

The intensities of the laser waves are always chosen such that the partial rates of the two direct terms $R_1$ and $R_2$ are practically equal. This setup is most favorable for studying interference because otherwise its contribution strength would be limited by the smaller of the direct terms.
It can be achieved by choosing a parameter $\zeta$, so that the two laser intensity parameters $\xi_i$ [compare \equref{laserwave}] are connected to the other wave's minimal photon number $\tilde{n}_i$:
\begin{equation}	
\xi_1 \approx \zeta^{\tilde{n}_2}
\quad\text{and}\quad
\xi_2 \approx \zeta^{\tilde{n}_1}.
\end{equation}
This leads to $R_1 \approx R_2 \approx \zeta^{2 \tilde{n}_1 \tilde{n}_2}$ because for the field intensities of interest here the direct rates depend on the respective intensity parameter like
\begin{equation}\equlab{rate-ampl}
R_i \sim \xi_i^{2 \tilde{n}_i},
\end{equation}
which can be understood from the fact that in leading order the rate contains the square of a Bessel function with an argument linear in the intensity parameter, corresponding to an $n$th-power dependence on that argument.
As \equref{rate-ampl} only gives the correct scaling, but does not contain the proportionality factor, which is indeed different for $R_1$ and $R_2$, small modifications are still necessary to obtain equal direct contributions. 
The final amplitudes can be found by starting from the above and calculating the direct terms once. From the comparison of their respective rates the according modification to the input amplitudes can be inferred.

In the following sections we show plots of angular-differential (partial) rates for various combinations of laser waves that feature prominent contributions from interference terms or, to contrast this, the lack thereof.
We investigate how the variation of the total photon energy and the relative phase between the two laser modes influences the interference terms and discuss these effects in nuclear rest frame and laboratory frame, the first to describe them in general and the latter for potential experimental application.

It should be noted that for the sake of clarity in the following figures, we do not depict those index combinations that would be allowed by energy conservation [described by the $\delta$-function in \equref{partcontrib}], but are several orders of magnitude smaller than the direct terms, as they represent higher orders in $\zeta$.
Also, for symmetry reasons there are always two interference terms in each plot contributing equally and thus overlaying each other. In all figures those two will be depicted by a single line, which has to be counted twice when summing up the terms.

Finally, it should be noted that for the results shown in the following sections we use $Z = 1$, assuming a proton beam target.
For higher nuclear charges the rates would be modified by an overall scaling factor of $Z^2$, as long as Coulomb corrections to our first-order treatment of the nuclear field are negligible (cf. \secref{totalenergy}). In this case our predictions of the influence of quantum interference can thus be inferred directly from the results for $Z = 1$.
%%
%It is worth emphasizing that from \equref{partcontrib} a nuclear-charge scaling of $\sim Z^2$ is found for the rates.
%%
%Using the alternative projectile available at the LHC, Pb with $Z = 82$, would thus result in an increase by almost four orders of magnitude.
%%
%However for heavier nuclei, the discussion of the necessity of Coulomb corrections found in \secref{totalenergy} would lead to the opposite conclusion, as they certainly would have to be taken into account for $Z \gtrsim 10$.

\subsection{Variation of the Minimal Photon Number in the Nuclear~Rest~Frame}

First, we shall present results for different laser pairs $(\tilde{n}_1, \tilde{n}_2)$ to find those candidates of laser wave combinations where the contributions originating from interference are most distinct.
Here we examine total photon energies of $\tilde{n}_1 \hbar \omega_1 = \tilde{n}_2 \hbar \omega_2 = 1.1 \unit{MeV}$ in the nuclear rest frame, with vanishing relative phase $\varphi$.
Photon energies of that magnitude would result, for instance, from colliding a nuclear beam with a Lorentz factor of $\gamma \approx 3000$, as the LHC provides in its current state, and an XUV laser beam of $\omega \sim 100 \unit{eV}$.
Alternatively, a much lower $\gamma$ of about $50$ could be used in conjunction with an X-ray laser beam  of $\omega \lesssim 10 \unit{keV}$.
As this consideration is confined to the nuclear rest frame, the  $\gamma$ of the nucleus is only used to reach the given total photon energy by a Lorentz boost of the laser frequencies. Its further influence will be discussed in \secref{labframe}, where the laboratory frame is studied.

Figure \ref{fig:const12} shows the angular-differential partial rates for laser pair $(1, 2)$. The emission angle $\theta$ is measured with respect to the laser propagation direction. 
% = \vec{e}_\text{z} = (0,0,1)
%
The positions of the emission peaks are generally at angles below $90\degree$ due to the momentum in forward direction imparted by the absorbed photons.
%as momentum conservation leads to a large momentum of the created particles in laser propagation direction.
%
The solid red line shows the contribution of the direct term stemming solely from the second mode, using two photons to create the pair.
Likewise, the long-dashed blue line is the contribution solely from the first mode, using one photon for pair creation.
Here the both terms stemming from interference (dotted violet lines) do not contribute significantly to the summed-up rate.
The same is true for laser pair $(1, 3)$ shown in \figref{const13}, where the above description applies accordingly.
Besides the two direct terms using the minimum number of photons, another direct term (dash-dotted magenta line) is visible in both plots, representing the next higher order in $\zeta$ for the second mode.
Two additional channels, one direct (dash--double-dotted cyan line) and one symmetrically mixed (double-dashed orange line), show a very small, yet non-negligible, contribution due to their even higher order in $\zeta$.

In addition, in both Figs. \ref{fig:const12} and \ref{fig:const13} the short-dashed green line depicts the symmetrically mixed channel $[1, 1, 1, 1]$, using one photon from each mode.
The most evident difference between $(1, 2)$ and $(1, 3)$ is the strength of this term, which is dominant in the first example and less pronounced, but not vanishing, in the latter.
In both cases it is expected to be suppressed because it is of a higher order in $\zeta$ compared to the direct term $[1, 1, 0, 0]$.
However, the strength of this term can be explained by the rather low total photon energy of $1.1 \unit{MeV}$, just above the pair-creation threshold, as it is known that the pair-creation rate vanishes at the threshold \cite{milstein}.
For total photon energies of the leading-order terms just above the threshold a term of higher order in $\zeta$, such as $[1, 1, 1, 1]$ here, has access to a substantially larger phase space for the emitted electron-positron pair, which can compensate its aforementioned suppression.
This effect vanishes for smaller intensities as the suppression for increased orders in $\zeta$ gets stronger as $\zeta$ becomes smaller.

\begin{figure}[t]
\begin{center}

\subfigure[\figlab{const12}
Laser pair $(1,2)$ with $\xi_1 = \sci{1.06}{-2}$ and $\xi_2 = 0.1$
]{%
\includegraphics[width=\columnwidth]{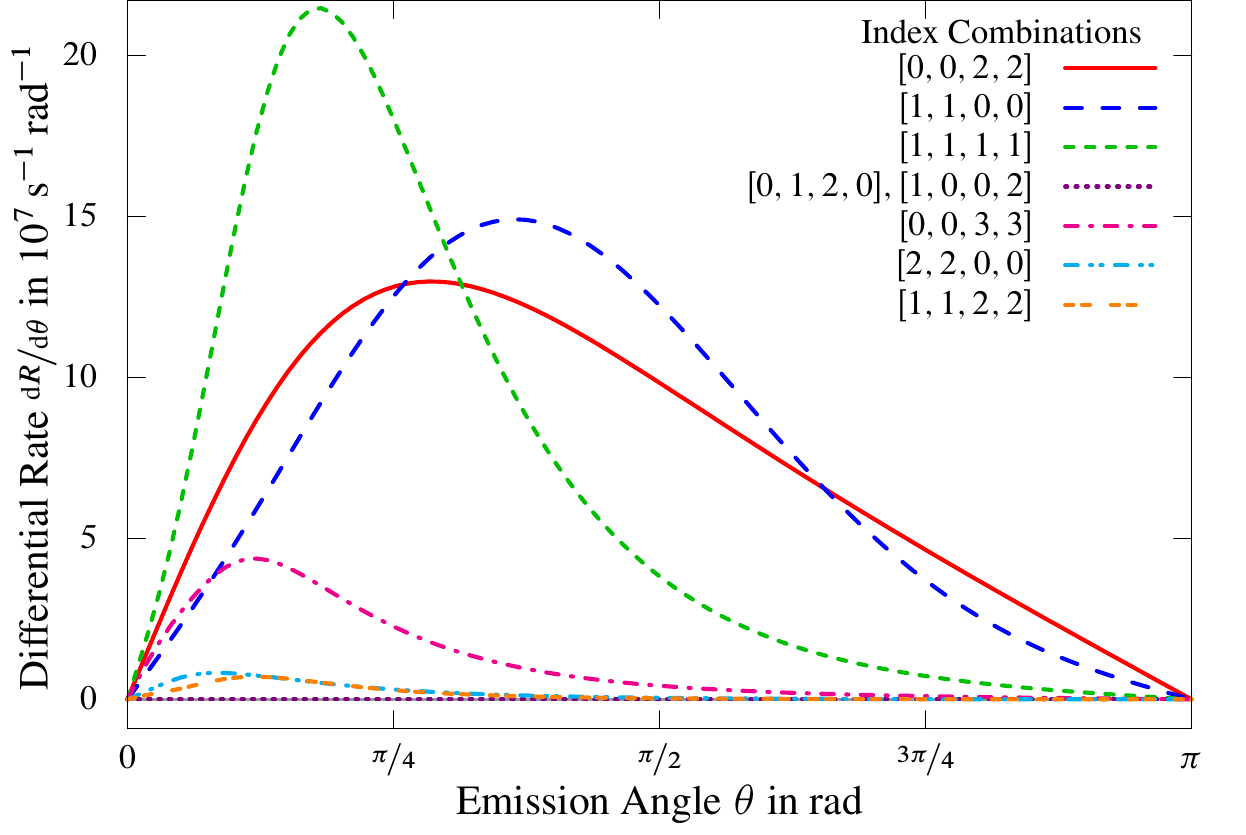}%
}

\subfigure[\figlab{const13}
Laser pair $(1,3)$ with $\xi_1 = \sci{3.59}{-4}$ and $\xi_2 = 0.1$
]{%
\includegraphics[width=\columnwidth]{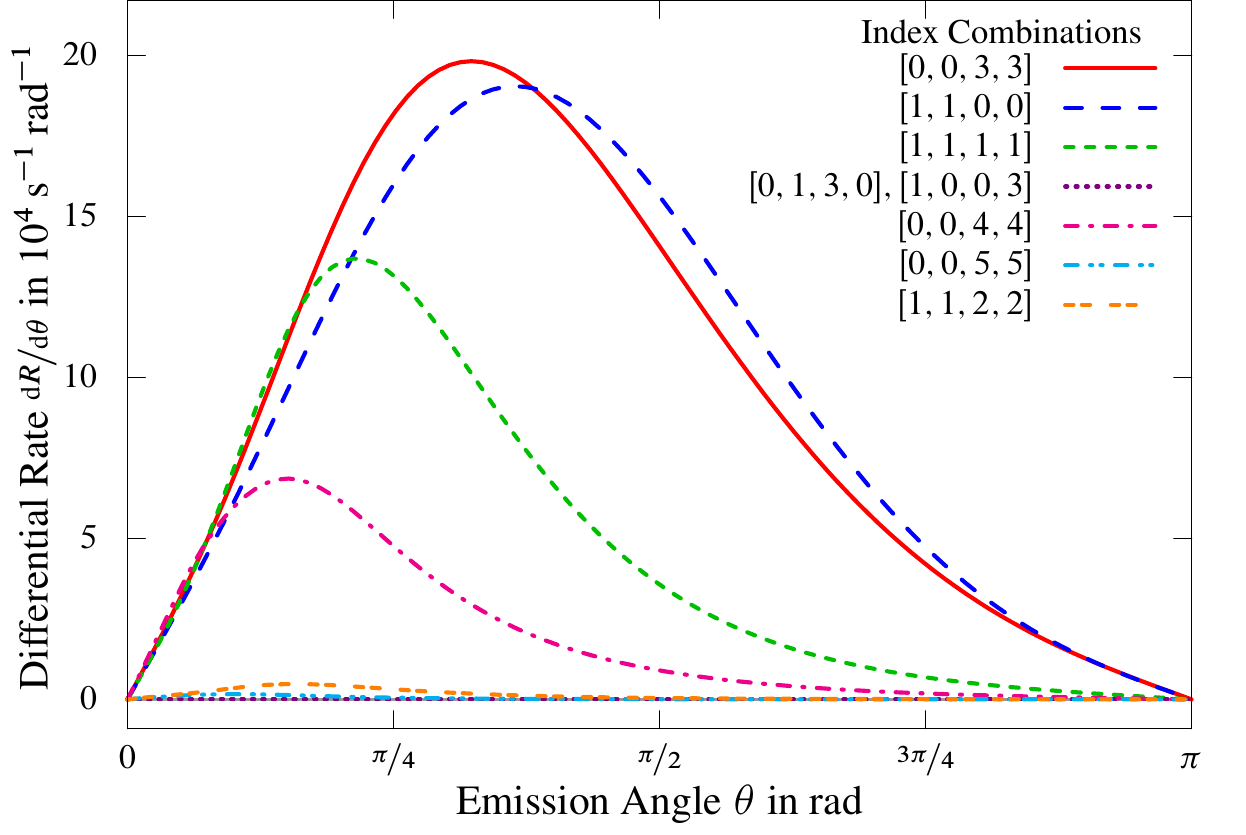}%
}

\caption{\figlab{const1}(Color online)~
Angular-differential partial rates for laser pairs $(1, 2)$ and $(1, 3)$ with total photon energy $1.1 \unit{MeV}$ in the nuclear rest frame --
The frequencies are chosen such that the total photon energy is reached for $\tilde{n}_1$ photons from the first mode and $\tilde{n}_2$ photons from the second mode.
The emission angle $\theta$ is measured with respect to the laser propagation direction.
The relative phase is set as $\varphi = 0$.
Both laser pairs show no significant contribution from an interference term.
}

\end{center}
\end{figure}

Contrasting these first two examples, the two laser pairs $(2, 4)$ and $(2, 6)$, as depicted in \figref{const2}, indeed show significant contribution from interference terms. For the chosen relative phase $\varphi = 0$ this contribution is destructive, i.e., leads to a decreased summed-up rate, or constructive, i.e., leads to an increased summed-up rate, for laser pairs $(2, 4)$ and $(2, 6)$, respectively.
In these two cases, the symmetrically mixed terms expectedly show contributions approximately as strong as the direct terms because their order in $\zeta$ is equal. Instead of taking all energy from a single mode, one half is taken from the first mode, the other half from the second mode.
Note that here intensities smaller than in the examples before are used; thus no contributions of higher orders in $\zeta$ are visible.

Qualitatively, the angular dependence of the interference terms may be understood as follows. The interference terms are sensitive to the relative phase between both laser modes. Variation of this phase implies that the total field vector of the laser changes (cf. \secref{relphase}). This leads to a redistribution of the emission angles, into which the created particles are ejected. Some angular regions might become more probable, whereas others become less probable. This redistribution is reflected, accordingly, by the angular dependence of the interference terms. Similar conclusions have been drawn regarding other strong-field processes, such as the two-color multiphoton ionization of atoms \cite{potvliege, schafer, telnov}.

Comparing our results to previous work on pair creation by a highly energetic photon \cite{narozhny}, where interference effects were found to be strongest for a frequency ratio of $3$, we note that in the present situation the laser pair with the smallest photon number for this ratio $(1,3)$ does not show a contribution from interference terms. However, we do see a strong contribution for the combination $(2, 6)$.
Furthermore, we do find laser pair $(2, 4)$ to show the highest absolute contribution from an interference term.  Again, we do not find any contribution for combination $(1,2)$, with lower photon number and identical frequency ratio.
The differences to the aforementioned study \cite{narozhny} may be attributable to the differing laser geometries investigated, as there parallel field vectors were considered.
In conclusion, we note that in our results interference seems to arise only when both minimal photon numbers $\tilde{n}_i$ are even.

As $(2, 4)$ shows the most prominent contribution from interference of all examined laser pairs, we shall use it as our main example for all studies in the following sections.

\begin{figure}[t]
\begin{center}

\subfigure[\figlab{const24}
Laser pair $(2, 4)$ with $\xi_1 = \sci{4.62}{-5}$ and $\xi_2 = 0.01$
]{%
\includegraphics[width=\columnwidth]{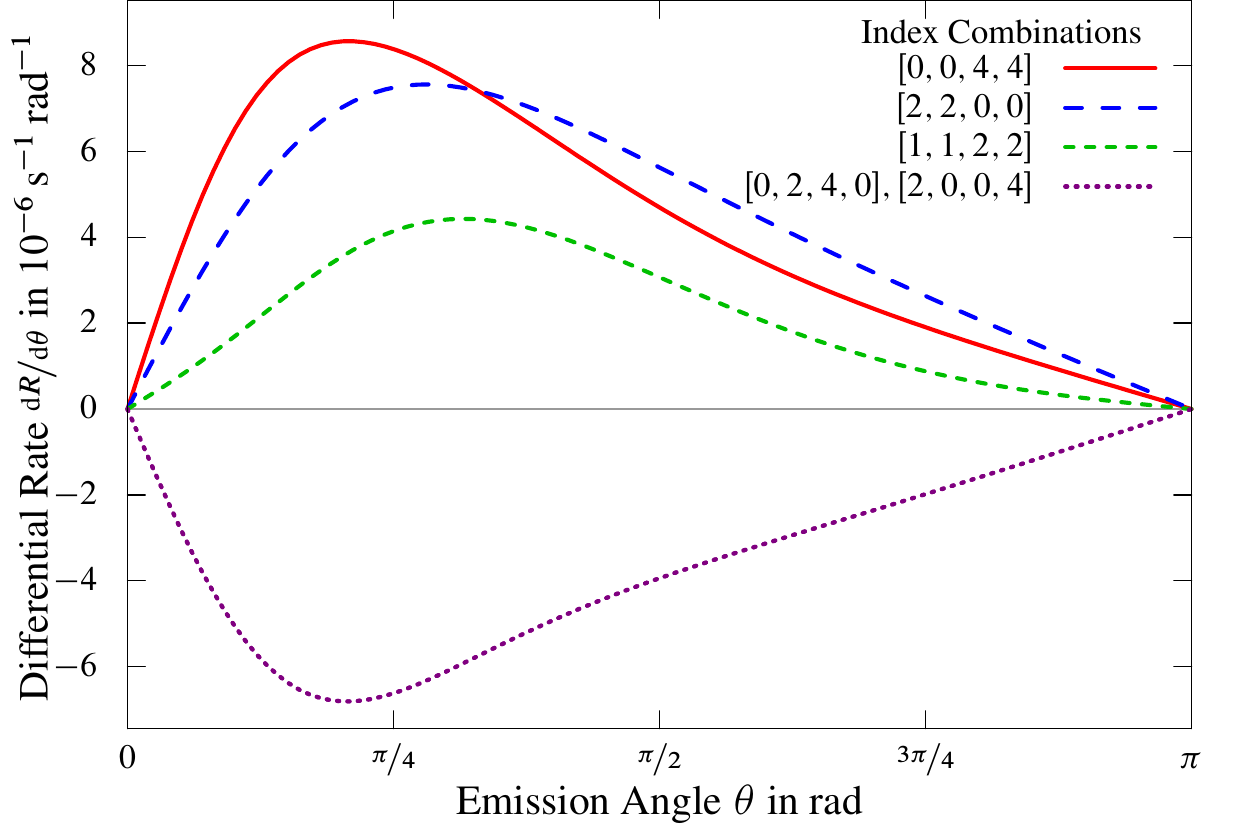}%
}

\subfigure[\figlab{const26}
Laser pair $(2, 6)$ with $\xi_1 = \sci{2.48}{-7}$ and $\xi_2 = 0.01$
]{%
\includegraphics[width=\columnwidth]{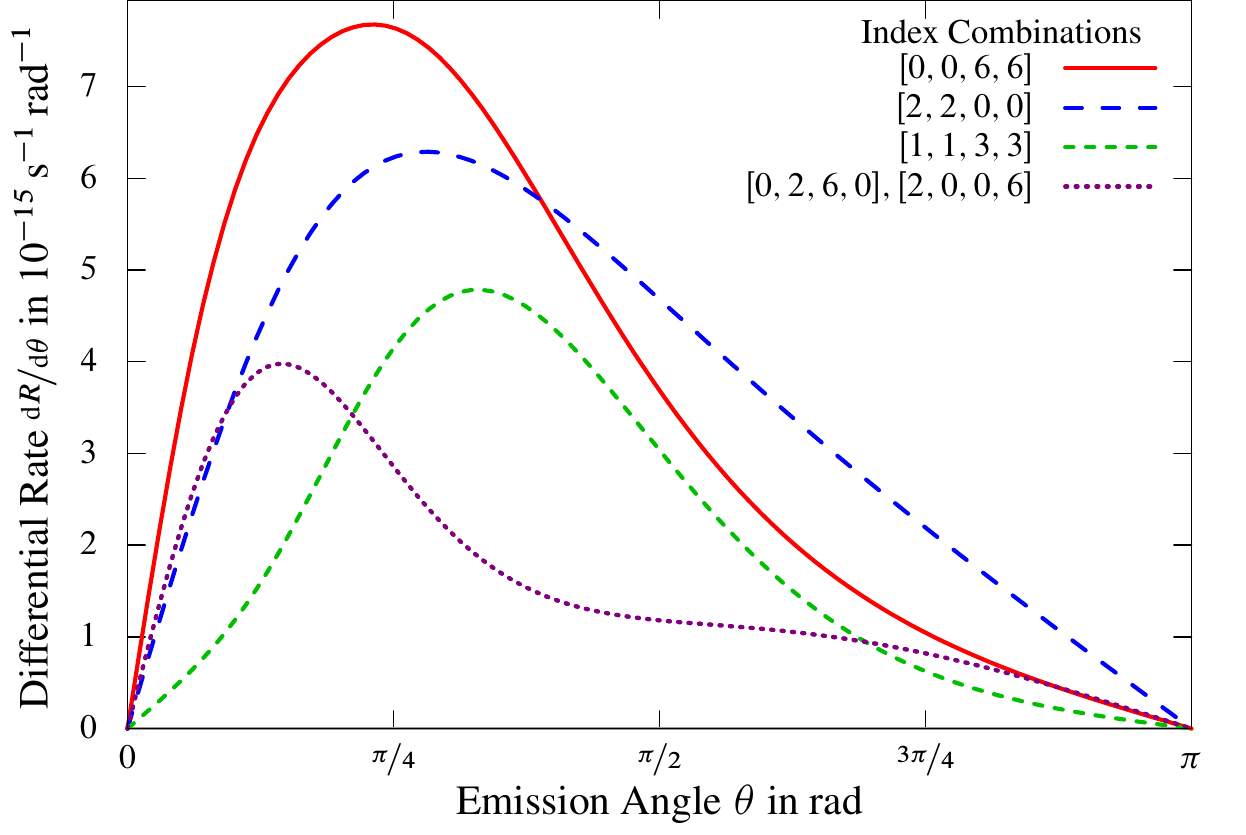}%
}

\caption{\figlab{const2}(Color online)~
Angular-differential partial rates for laser pairs $(2, 4)$ and $(2, 6)$ --
%
%If not indicated otherwise, the parameters are the same as in \figref{const1}.
%
These two laser pairs show a large contribution from interference terms. For the relative phase $\varphi$ set to zero, interference is destructive for $(2, 4)$ and constructive for $(2, 6)$.
}

\end{center}
\end{figure}

\subsection{Variation of the Total Photon Energy in the Nuclear~Rest~Frame} \seclab{totalenergy}

For a given laser pair, here $(2, 4)$, a variation of the total photon energy in the nuclear rest frame shows different modifications for the different terms in the sum in \equref{amplitude}.
This is depicted in \figref{24energ}, where the angular-differential partial rates are shown for total photon energies of $1.25$ and $1.35 \unit{MeV}$.

\begin{figure}[t]
\begin{center}
\subfigure[\figlab{24energ125}
Total photon energy $1.25 \unit{MeV}$, $\xi_1 = \sci{5.07}{-5}$, $\xi_2 = 0.01$
]{%
\includegraphics[width=\columnwidth]{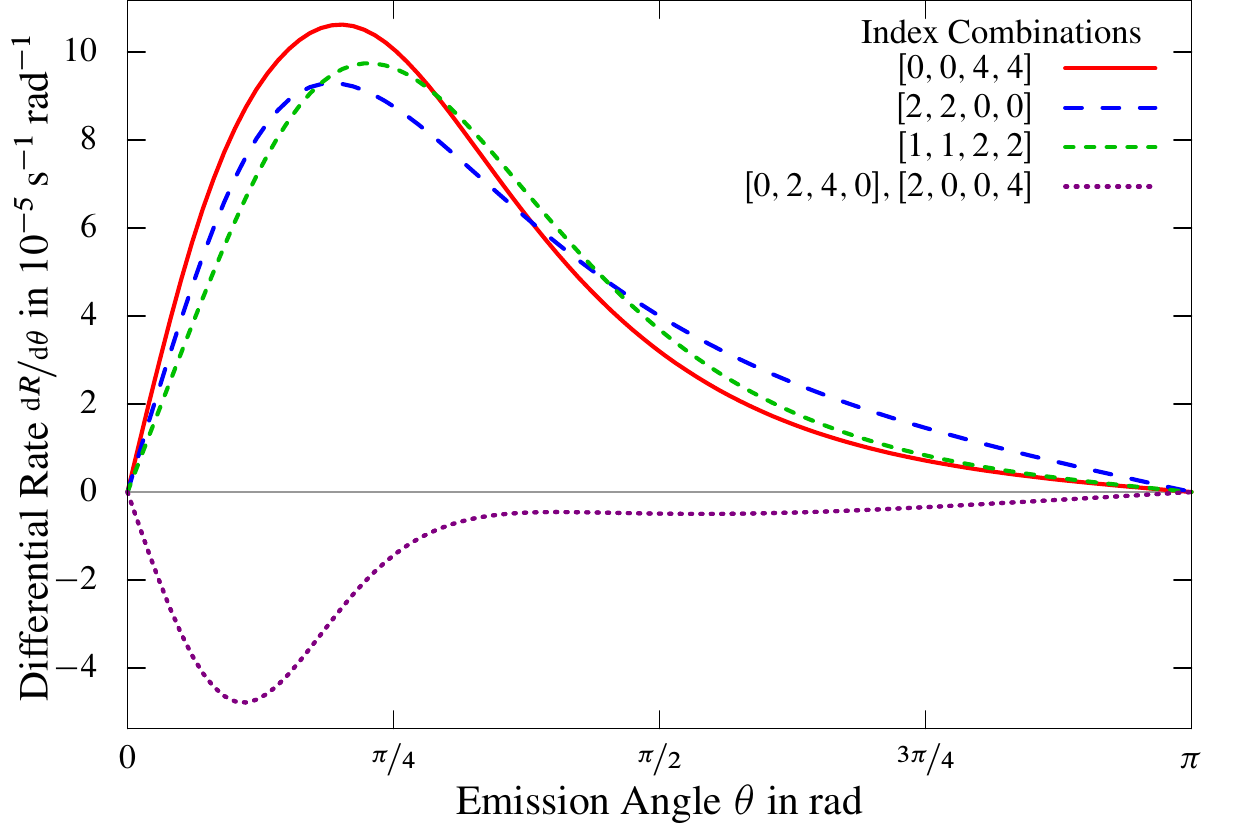}%
}
\subfigure[\figlab{24energ135}
Total photon energy $1.35 \unit{MeV}$, $\xi_1 = \sci{5.38}{-5}$,  $\xi_2 = 0.01$
]{%
\includegraphics[width=\columnwidth]{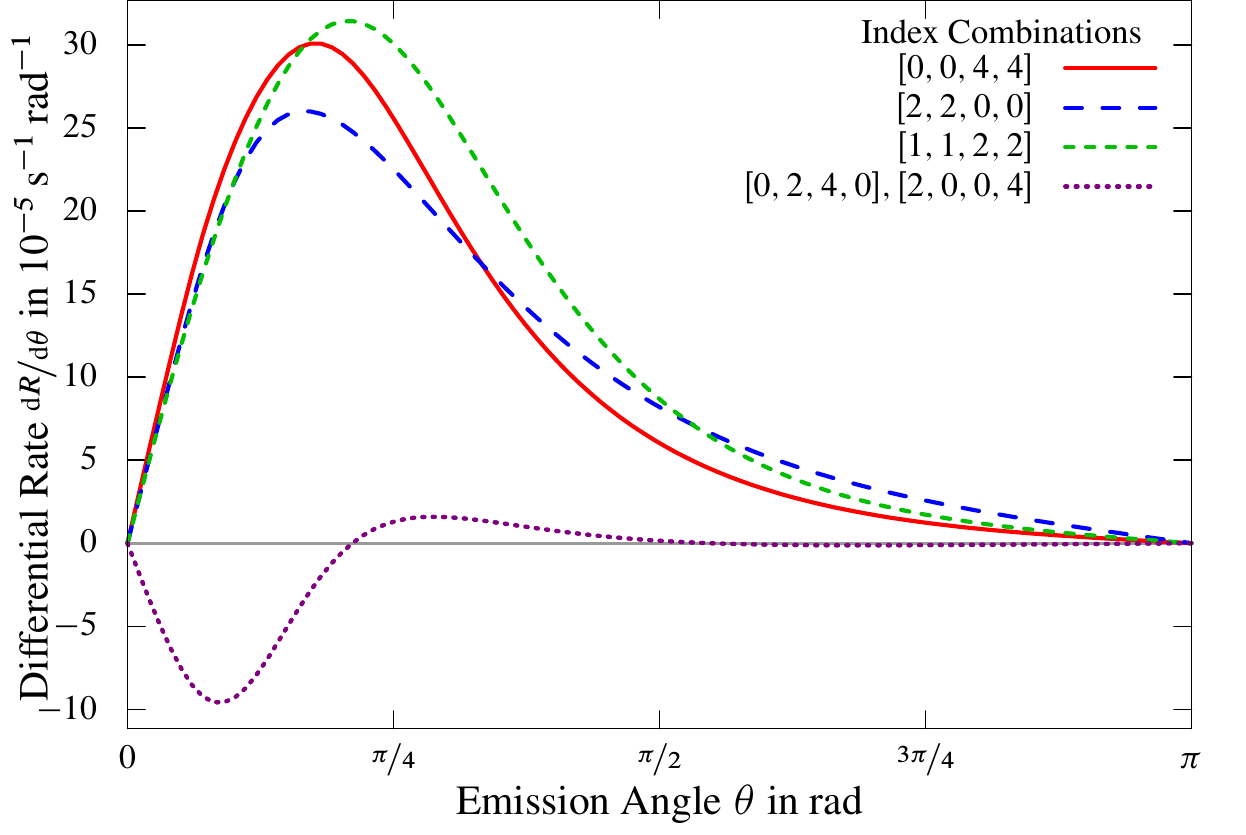}%
}
\caption{\figlab{24energ}(Color online)~
Variation of the total photon energy of the angular-differential partial rates for laser pair $(2, 4)$ --
As the total photon energy increases, the relative contribution from interference decreases. Additionally, the shape of the interference terms changes, leading to a zero crossing in \figref{24energ135}.
The intensity parameter $\xi_1$ has been adjusted to maintain equally contributing direct terms.
}
\end{center}
\end{figure}

While all terms tend to be narrowed to the lower part of the angular spectrum for higher energies, the interference terms (violet dotted line) additionally show clear changes in their shape. Particularly, the zero crossing for $1.35 \unit{MeV}$ [\figref{24energ135}] is an interesting feature.
%, and will be discussed later.

Furthermore, the relative contribution of the symmetrically mixed term (short-dashed green line) grows with increasing total photon energy, making it dominant for the highest depicted energy.
It should also be noted that for even higher total photon energies, i.e., those larger than $1.363 \unit{MeV}$, a new direct reaction channel, $[0, 0, 3, 3]$, opens and dominates the rate from $1.375 \unit{MeV}$ onwards. As this is not relevant for our investigation, we do not give figures for energies higher than $1.35 \unit{MeV}$.
We can conclude that the interference is more pronounced close to the pair-creation threshold, while the change in their curve shape for higher energies might lead to interesting effects.

Since interference is most visible for total photon energies that are rather close to the threshold, the question arises whether Coulomb corrections %\cite{coulombwave1, *coulombwave2} 
to the outgoing particles are relevant. 
However, even for the lowest considered total photon energy of $1.1 \unit{MeV}$ the created particles are fast enough to leave the vicinity of the nucleus without being deflected substantially. This corresponds to their Sommerfeld parameter \cite{LL3} being small:
%
%On a side note, in our case it is permitted to neglect so-called Coulomb corrections, which would include the attraction or repulsion of the created particles (electron and positron, respectively), due to the Coulombic field of the nucleus \cite{coulombwave1, coulombwave2}.
%
%These have to be considered if the particles are slow enough to feel the nuclear field. But even for our case of low total photon energy, the kinetic energy is sufficiently large for the particles to leave the vicinity of the nucleus without being deflected.
%
%To find a measure for the influence of the nuclear field on a nearby electron or positron, one can compare the squared modulus of the wavefunction for a Coulomb field with that of the free particle counterpart at the coordinate origin, i.e., at the position of the nucleus \cite{LL3}. If no deviation is found, the assumption that the influence of the nuclear Coulomb field on the trajectory of the particles can be neglected, is justified.
%
%Starting from \cite{LL3}, it can be shown that this is the case if
%
$$
\frac{Z \alpha}{\beta_{\pm}} \ll 1.
$$
Here $Z$ is the nuclear charge, $\alpha \approx \nicefrac{1}{137}$ the fine structure constant, and $\beta_{\pm} = \nicefrac{v_{\pm}}{c}$ the respective created particle's velocity in units of the speed of light.
This quantity is indeed very small 
for typical kinetic energies of about $40 \unit{keV}$, corresponding to velocities of $\beta_{\pm} \approx 0.4$, and small $Z$. Particularly, for $Z = 1$, as we consider here, we obtain
$
\nicefrac{Z \alpha}{\beta_{\pm}} \approx 0.02.
$
Consequently, Coulomb corrections are of minor importance for the total photon energies under consideration here.

\subsection{Variation of the Relative Phase in the Nuclear~Rest~Frame} \seclab{relphase}

To study the influence of the relative phase $\varphi$ -- we recall that $\varphi = \varphi_1$ and $\varphi_2 = 0$ -- between the two laser modes of laser pair $(2, 4)$, it is helpful to first consider a few sub-sums of the sum in \equref{amplitude}. This is shown in \figref{ph-subsum} for two total photon energies in the nuclear rest frame. Figure \ref{fig:ph-subsum-110} treats an energy of $1.1 \unit{MeV}$, just above the threshold of $2 m_* c^2$. Figure \ref{fig:ph-subsum-135} treats $1.35 \unit{MeV}$, just below the energy at which the additional reaction channel opens, as discussed in the section before.

In these two Figs. \ref{fig:ph-subsum-110} and \ref{fig:ph-subsum-135} the individual terms of the angular-differential rate are depicted as before in \figref{const24}, retaining all parameters except the relative phase $\varphi$, which is set to $\nicefrac{\pi}{2}$ instead. As we see, the interference term has changed sign and thus has become constructive.

The individual terms are compared with the sum of all terms (dash-dotted magenta line), the sum of the direct terms $[0,0,4,4]$ and $[2,2,0,0]$ (dash--double-dotted cyan line), the sum of the phase-independent contributions (double-dashed orange line), which are the latter two and the symmetrically mixed term $[1,1,2,2]$, and the sum of the two interference terms, $[0,2,4,0]$ and $[2,0,0,4]$ (double-dash--dotted yellow line).

The comparison shows that for the lower total photon energy the sum of the direct terms amounts to about $50\%$ of the summed-up rate, the symmetrically mixed term adds up to about $10\%$, and the sum of the interference terms contributes about $40\%$.
For the higher energy the interference terms are less pronounced, yielding only about $10\%$, with the phase-independent terms giving the other $90\%$ (which consists of three similar shares stemming from the two direct terms and the symmetrically mixed term).

\begin{figure}[t]
\begin{center}

\subfigure[\figlab{ph-subsum-110}
Total photon energy of $1.1 \unit{MeV}$
]{%
\includegraphics[width=\columnwidth]{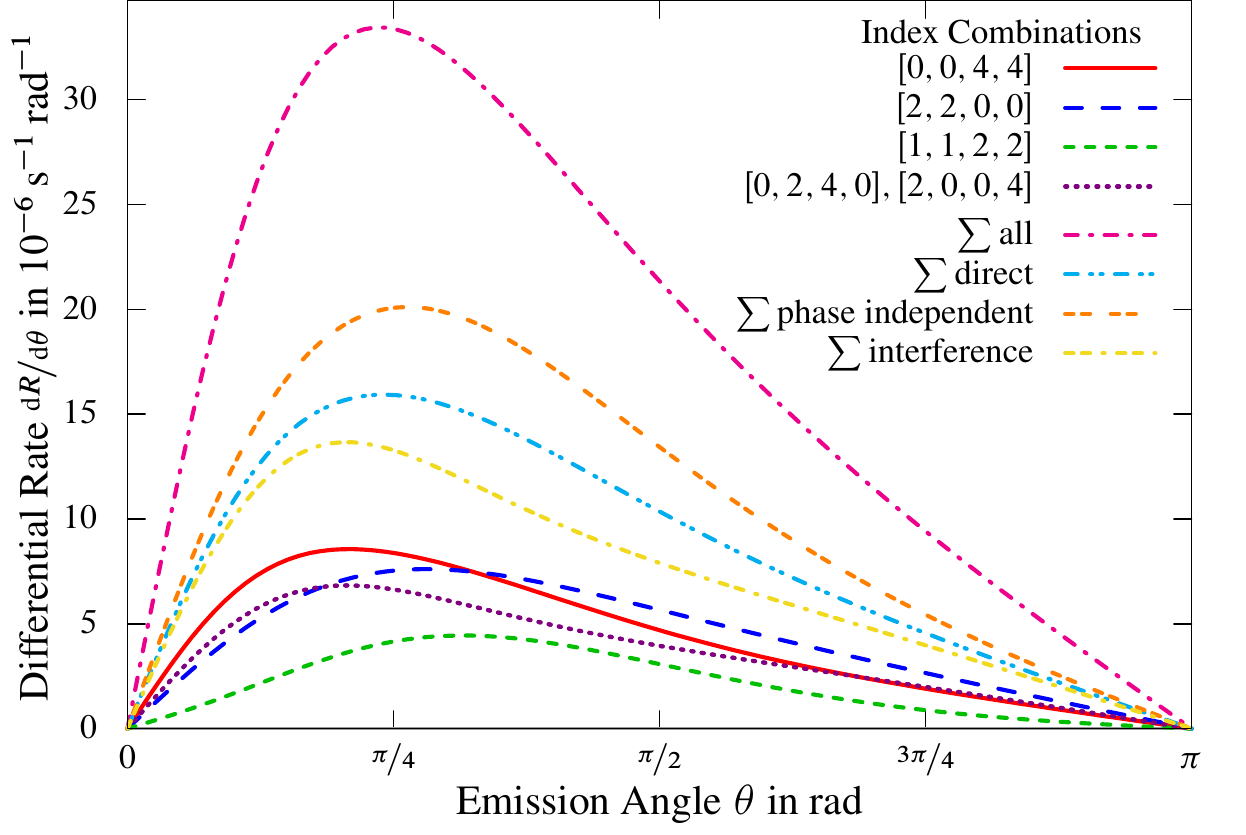}%
}

\subfigure[\figlab{ph-subsum-135}
Total photon energy of $1.35 \unit{MeV}$
]{%
\includegraphics[width=\columnwidth]{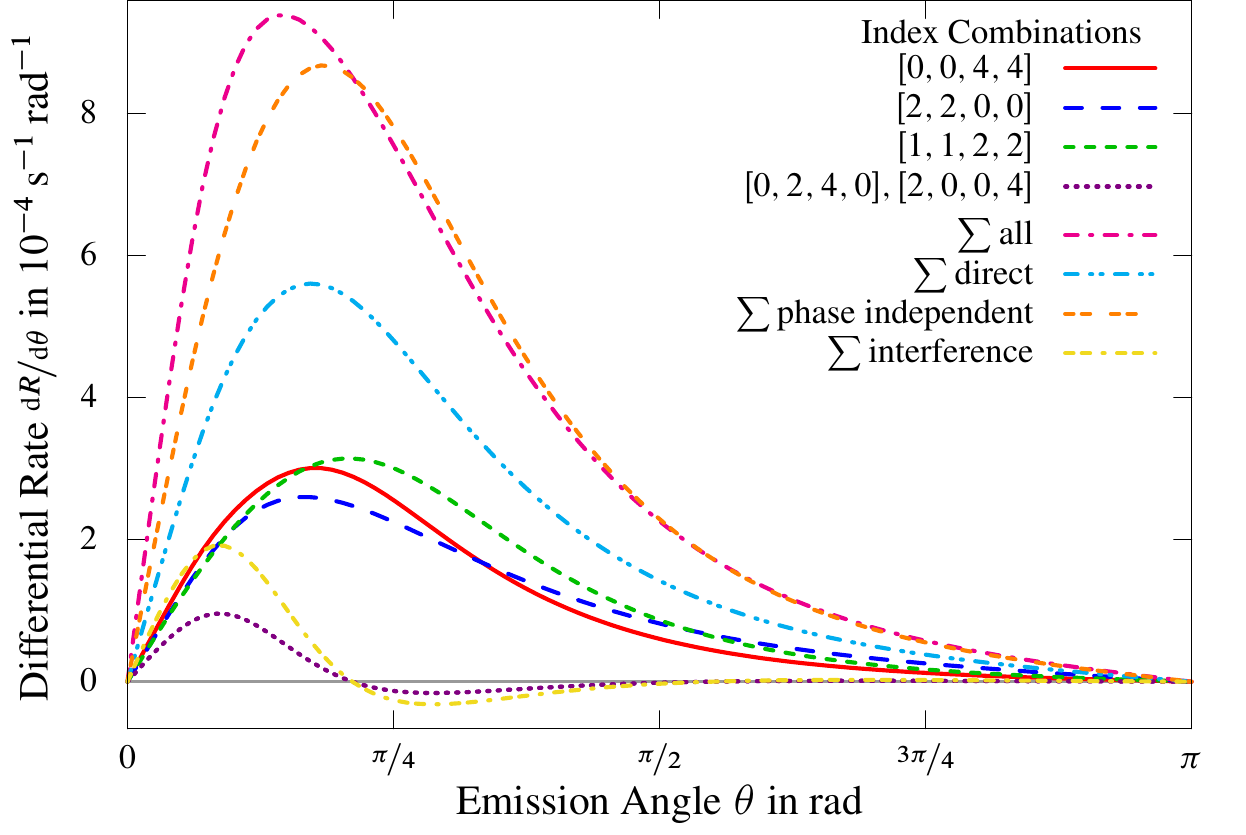}%
}

\caption{\figlab{ph-subsum}(Color online)~
Individual terms and several sub-sums of the angular-differential partial rates of laser pair $(2, 4)$ for two photon energies --
Here \quot{$\sum$\,\text{all}} is the sum over all individual terms, \quot{$\sum$\,\text{direct}} is the sum of the two terms stemming solely from the first or the second laser mode, \quot{$\sum$\,\text{phase independent}} is the latter plus the symmetrically mixed term, and \quot{$\sum$\,\text{interference}} is the sum of the two interference terms. The relative phase is set as $\varphi = \nicefrac{\pi}{2}$. Otherwise the parameters are the same as in \figref{const24} and \figref{24energ135}.
}

\end{center}
\end{figure}

For the same two energies \figref{ph-sumint} shows a comparison of several relative phases for the sum of all terms (top panels) and for the sum of just the two interference terms (bottom panels).
In the top panels the sum of the direct terms is given as a height reference.

\begin{figure}[t]
\begin{center}

\subfigure[\figlab{ph-sumint-110}
Total photon energy of $1.1 \unit{MeV}$
]{%
\includegraphics[width=\columnwidth]{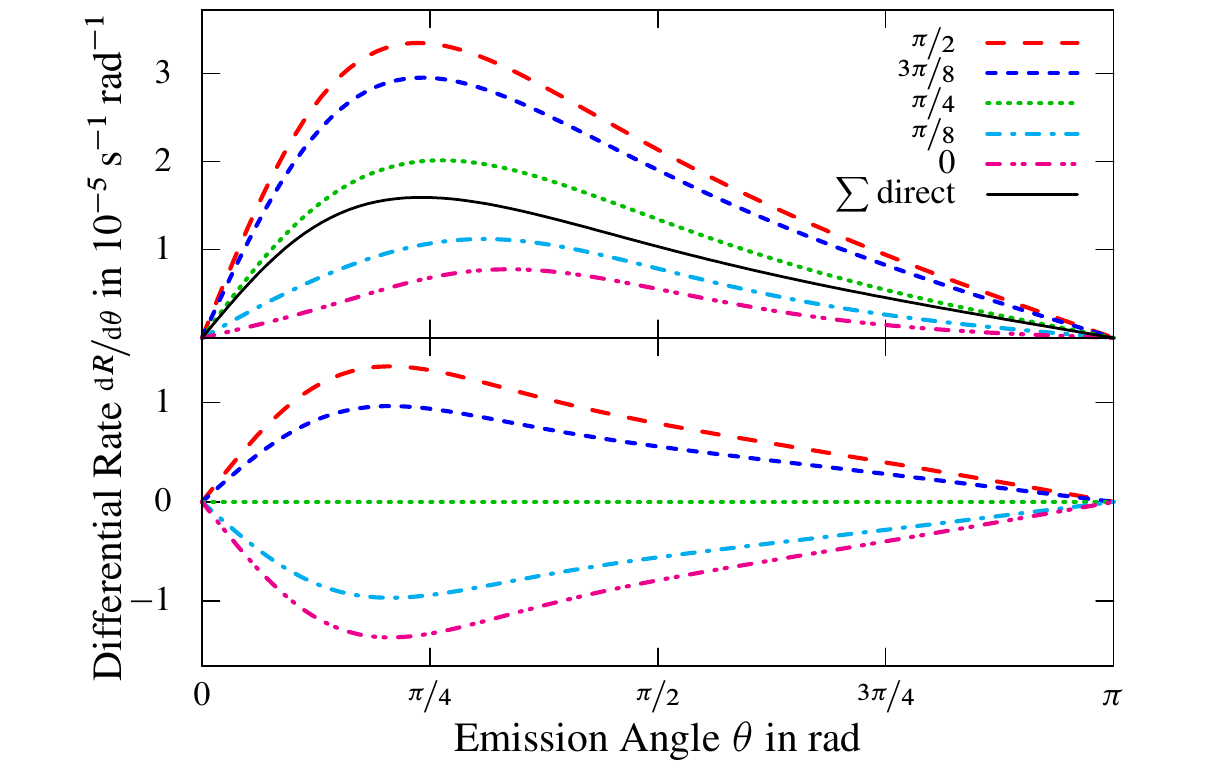}%
}

\subfigure[\figlab{ph-sumint-135}
Total photon energy of $1.35 \unit{MeV}$
]{%
\includegraphics[width=\columnwidth]{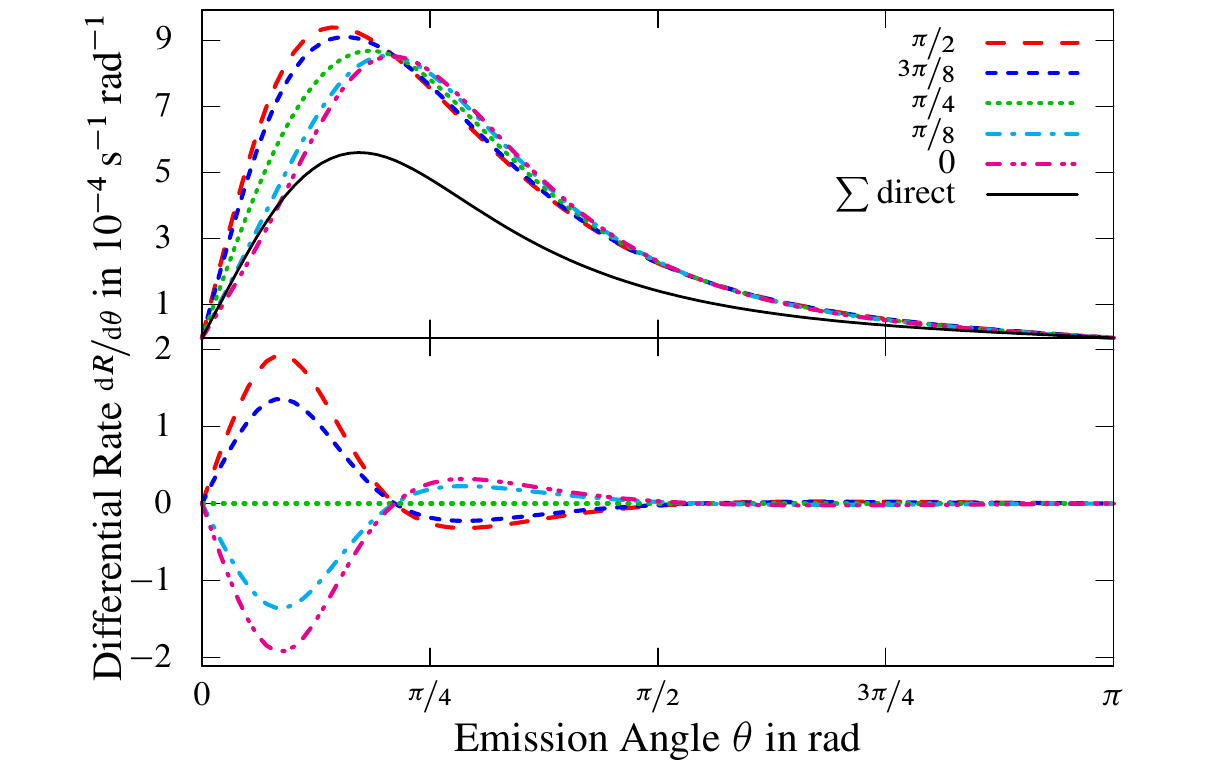}%
}

\caption{\figlab{ph-sumint}(Color online)~
Phase variation in the sums of all terms (top panels) and the interference contributions (bottom panels) of the angular-differential partial rates of laser pair $(2, 4)$ for two energies. The sum of the direct terms as in \figref{ph-subsum} is given for reference --
For laser pair $(2, 4)$ the interference contribution is maximal (maximally constructive) for the relative phase $\varphi = \nicefrac{\pi}{2}$, is minimal (maximally destructive) for $\varphi = 0$, and vanishes for $\varphi = \nicefrac{\pi}{4}$.
The other parameters are the same as in \figref{const24} and \figref{24energ135}, respectively.
}

\end{center}
\end{figure}

We find the interference terms of laser pair $(2, 4)$ to have a sinusoidal behavior in the relative phase $\varphi$, particularly they show a periodicity of $\nicefrac{2 \pi}{\tilde{n}_1} = \pi$. As mentioned earlier, for vanishing $\varphi$ the interference terms contribute destructively to the rate of this laser pair. By tuning the relative phase to $\nicefrac{\pi}{2}$ this contribution becomes maximally constructive.
Any phase in between will lead to a decrease in the absolute interference contribution, with a phase of $\nicefrac{\pi}{4}$ removing it completely.

In the two top panels in \figref{ph-sumint} the influence of phase variation on the angular-differential rate is shown. Besides the obvious increase or decrease in height (for constructive and destructive interferences, respectively), angular shifts of the peak position can be observed.
This means phase variation between the two laser waves could have a use not only in maximizing the pair-production rates but also in steering the direction of the peak emission.

The height change is prominently illustrated by \figref{ph-sumint-110} for $1.1 \unit{MeV}$, where a factor of approximatively $5$ can be seen between the summed-up rate for $\varphi = 0$ and $\nicefrac{\pi}{2}$.
While the shift in the peak position is also visible for the lower energy, it can be seen clearly in \figref{ph-sumint-135} for $1.35 \unit{MeV}$, where it is pronounced due to the aforementioned zero crossing of the interference terms.
This means that for the lower energy the phase variation moves the summed-up rate from below to above the sum of the direct terms and that for the higher energy the summed-up rate always stays above the sum of the direct terms but moves from larger to smaller angles.

\begin{figure}[t]
\begin{center}
\includegraphics[width=\columnwidth]{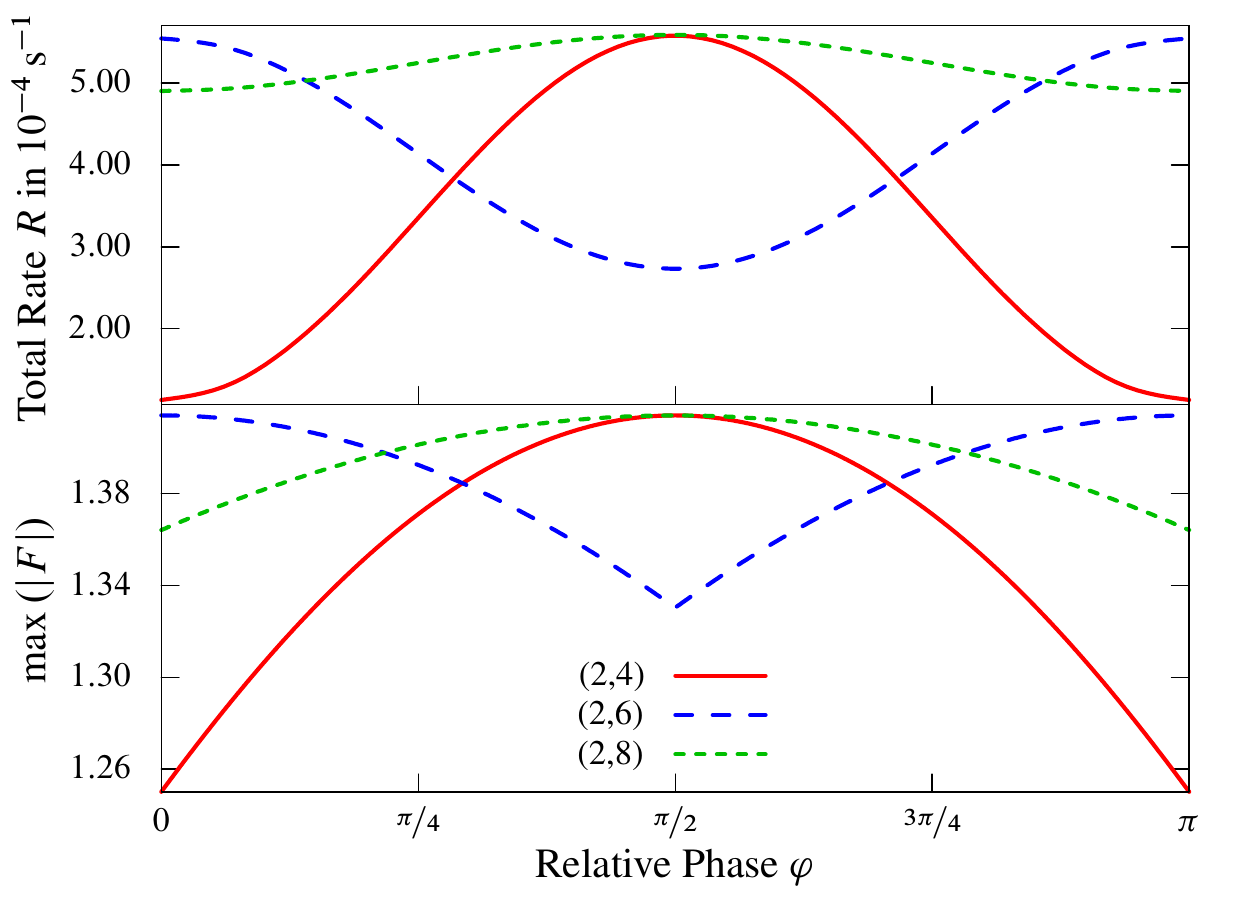}%
\caption{\figlab{ph-meas}(Color online)~
Comparison of (top) the scaled phase variation in the total pair-production rates and (bottom) the measure defined by the absolute maximum of the sum of the normalized electric fields from \equref{normelectricfield} for several laser pair combinations --
Line types and colors are identical for both panels.
As before $\varphi = \varphi_1$ and $\varphi_2 = 0$.
To achieve better comparability the total rates have been scaled by a factor of $\sci{1.45}{9}$ for laser pair $(2, 6)$ and by $\sci{1.35}{18}$ for $(2,8)$.
}
\end{center}
\end{figure}

It should be stressed again that we arbitrarily decided to vary $\varphi = \varphi_1$. Had we used $\varphi_2$ instead, we would have seen a periodicity of $\nicefrac{2 \pi}{\tilde{n}_2}$, as it depends on the laser wave frequency and thus on the respective minimal number of photons $\tilde{n}_i$.
An intuitive explanation for the found periodicity of $\nicefrac{2 \pi}{\tilde{n}_i}$ in the relative phases $\varphi_i$ can be gained by studying the electric fields associated with the laser waves $A_i$ given in \equref{laserwave}:
\begin{equation}
\vec{E}_i = - \frac{1}{c} \frac{\partial \vec{A}_i}{\partial t}
= \frac{\vec{a}_i \omega_i}{c} \sin(\eta_i + \varphi_i)
\quad
\left(i=1 ~\text{or}~ 2\right).
\end{equation}
Due to the perpendicular laser-wave field-vectors $\vec{a}_i$, the square of the total electric field $\vec{E} = \vec{E}_1 + \vec{E}_2$ has the form
\begin{equation}
E^2 (ct-z)
= \sum_{i=1}^{2} E_i^2 
= \sum_{i=1}^{2}
\frac{\vec{a}_i^2 \omega_i^2}{c^2} \sin^2(\eta_i + \varphi_i),
\end{equation}
which can be used to calculate a measure for the phase dependence. Recall that $\nicefrac{\eta_i}{\omega_i} = t - \nicefrac{z}{c}$. Instead of using $E_i^2$ directly, we normalize the amplitudes
\begin{equation} \equlab{normelectricfield}
F^2 (ct-z) = 
\sum_{i=1}^{2} \frac{c^2}{\vec{a}_i^2 \omega_i^2} E_i^2
 = \sum_{i=1}^{2}
 \sin^2(\eta_i + \varphi_i),
\end{equation}
and take the modulus of the hence defined $F$. Its maximum absolute value $\max(\abs{F})$ is given in \figref{ph-meas} as function of the relative phase $\varphi = \varphi_1$ (bottom panel), showing the same periodicity as the total pair-creation rate (top panel). Thus we can conclude that the phase dependence of the total rate is directly retained from the electric fields of the input laser waves.
In contrast, interference patterns in the spectra of created particles are connected to the phase dependence of the vector potential, as has been shown in an earlier study \cite{krajewska-symm}.

\subsection{Variation of the Relative Phase in the Laboratory~Frame} \seclab{labframe}

Up until now we have presented results found in the nuclear rest frame; now we shall transform these findings to the laboratory frame to provide a comparison for potential experimental results.

As before, the laser wave combination $(2, 4)$ shall be considered, again with total photon energies of $1.1$ and $1.35 \unit{MeV}$ in the nuclear rest frame. As mentioned earlier, the Lorentz factor $\gamma$ used to reach these photon energies is irrelevant as long as the calculation takes place in the nuclear rest frame only. This is obviously not true for calculations in the laboratory frame, as $\gamma$ influences the transformation from the nuclear rest frame to laboratory.
Particularly, the photon energy sees a boost by a factor of  $(1+\beta) \gamma$, with $\beta = \sqrt{1-\nicefrac{1}{\gamma^2}}$.
For the following results in the laboratory frame we consider $\gamma = 50$, and thus, for instance, for the total photon energies of $1.1 \unit{MeV}$
\begin{equation}
\omega_1 = 5.50 \unit{keV}
\quad\text{and}\quad
\omega_2 = 2.75 \unit{keV}.
\end{equation}

Figures \ref{fig:boost-ph-subsum} and \ref{fig:boost-ph-sumint} show Lorentz-transformed versions of Figs. \ref{fig:ph-subsum} and \ref{fig:ph-sumint}, respectively.
Here the positions of the emission peaks are at angles just below $180\degree$, as the Lorentz-boost leads to the nuclear propagation direction being preferential for the created particles.
Comparing the shape of the individual graphs, one can observe that the direct terms are much less differentiated for both energies, while for the higher energy the interference terms show a more balanced ratio of negative and positive amplitude.

In the nuclear rest frame (\figref{ph-subsum}) the direct term $[0,0,4,4]$ is larger than the direct term $[2,2,0,0]$ for $\theta \lesssim 1$ and vice versa for $\theta \gtrsim 1$.
This difference vanishes almost completely in the laboratory frame [\figref{boost-ph-subsum}], where these two terms practically overlap each other.
For the total photon energy $1.35 \unit{MeV}$ in the nuclear rest frame (\figref{ph-subsum-135}) the interference terms show a much stronger absolute contribution for emission angles $\theta \lesssim 0.66$ than for larger angles.
This is again different in the laboratory frame [\figref{boost-ph-subsum-135}], where the contribution from the left and the right of the zero crossing at $\theta \approx 3.12$ is of similar absolute peak height.

In the phase variation shown in \figref{boost-ph-sumint} a modification of both peak position and height is again clearly visible, although here the latter is over a much smaller angular range, as the whole spectrum is contracted to a narrower angular distribution.
Another notable difference between the two reference frames is the direction of the aforementioned peak position shift. While in the nuclear rest frame the shift is towards smaller emission angles for a growing relative phase $\varphi$, the direction is reversed in the laboratory frame. This behavior is mirrored by the change of sign of the interference terms for the higher total photon energy [cf. the bottom panels of Figs. \ref{fig:ph-sumint-135} and \ref{fig:boost-ph-sumint-135}].
The effect can be attributed to the nature of the Lorentz transformation, which maps small angles in the nuclear rest frame to large ones in the laboratory frame and vice versa.

\begin{figure}[t]
\begin{center}

\subfigure[\figlab{boost-ph-subsum-110}
Total photon energy (in nuclear rest frame) of $1.1 \unit{MeV}$
]{%
\includegraphics[width=\columnwidth]{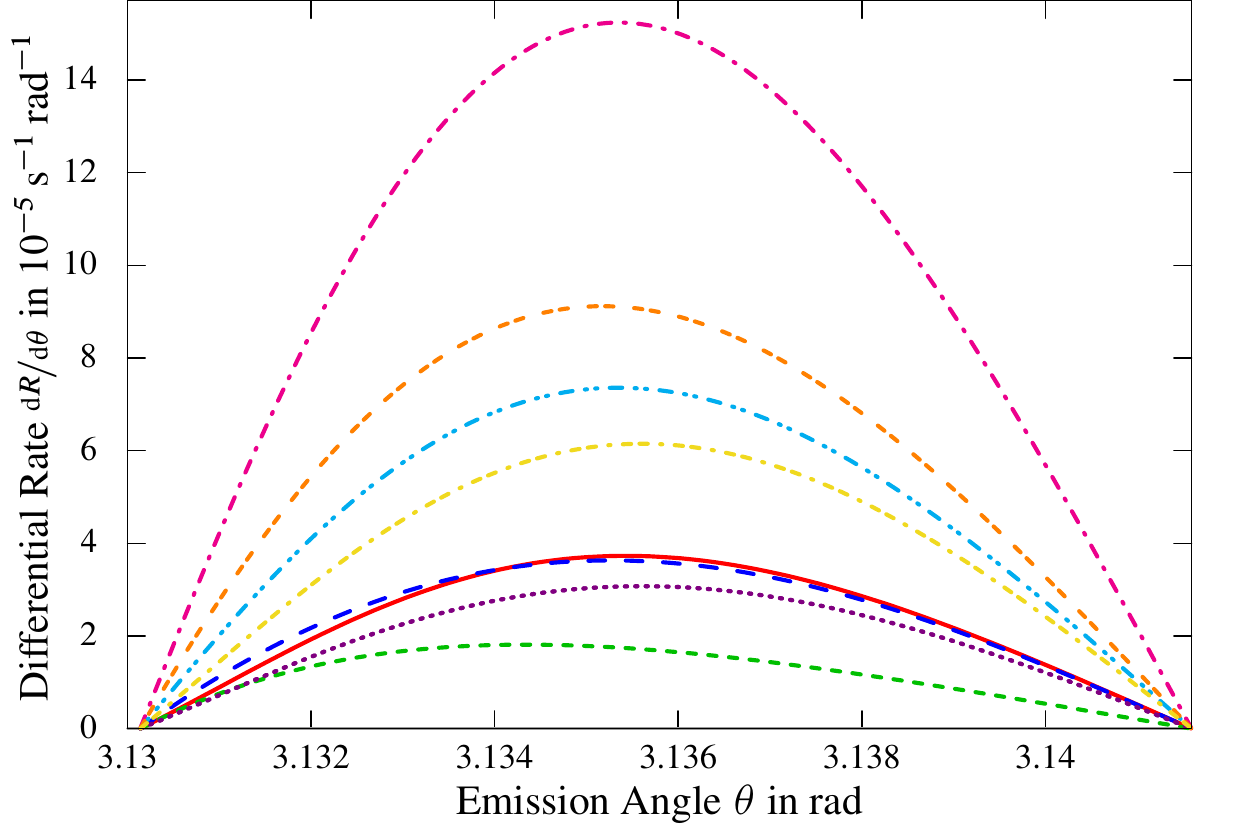}%
}

\subfigure[\figlab{boost-ph-subsum-135}
Total photon energy (in nuclear rest frame) of $1.35 \unit{MeV}$
]{%
\includegraphics[width=\columnwidth]{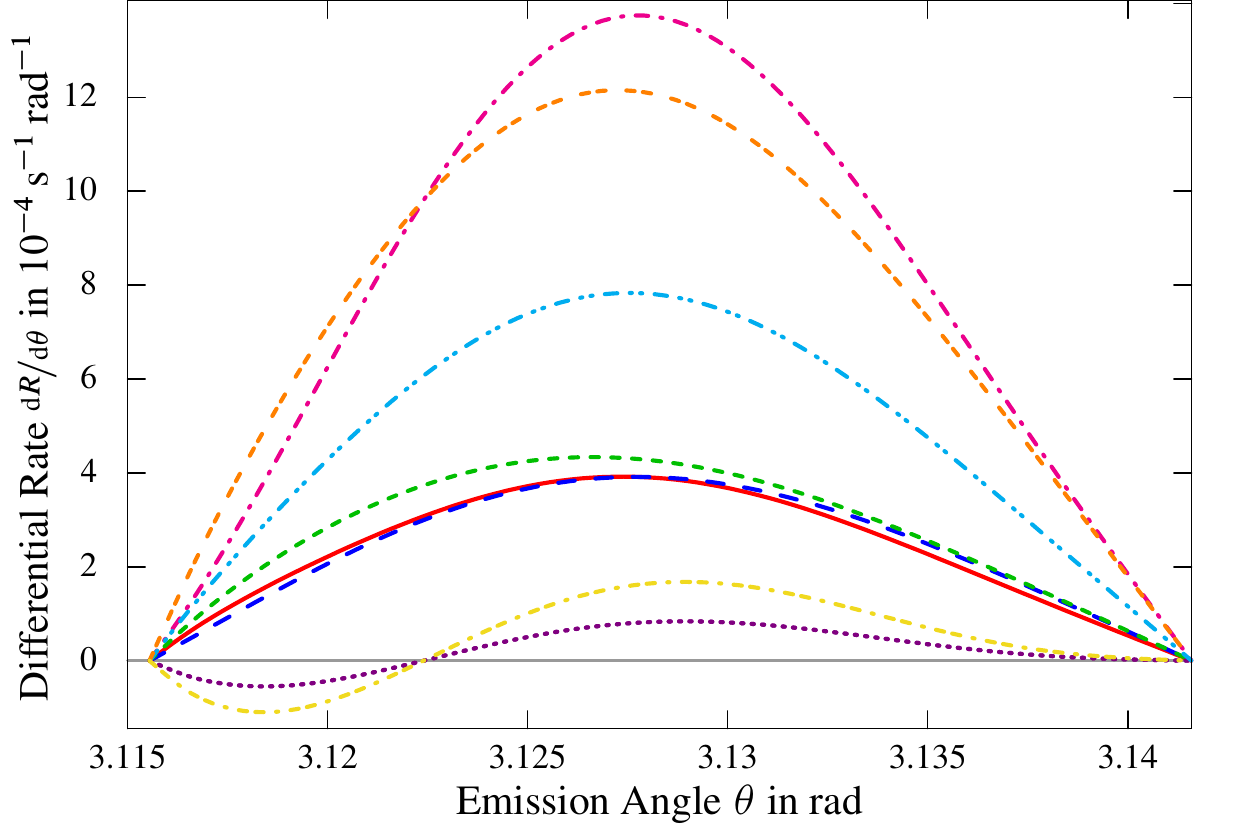}%
}

\caption{\figlab{boost-ph-subsum}(Color online)~
Individual terms and several sub-sums of the angular-differential partial rates of laser pair $(2, 4)$ for two photon energies transformed with a Lorentz factor $\gamma$ of $50$ to the laboratory frame --
Notation, legend, and parameters are identical to those in \figref{ph-subsum}.
}

\end{center}
\end{figure}

\begin{figure}[t]
\begin{center}

\subfigure[\figlab{boost-ph-sumint-110}
Total photon energy (in nuclear rest frame) of $1.1 \unit{MeV}$
]{%
\includegraphics[width=\columnwidth]{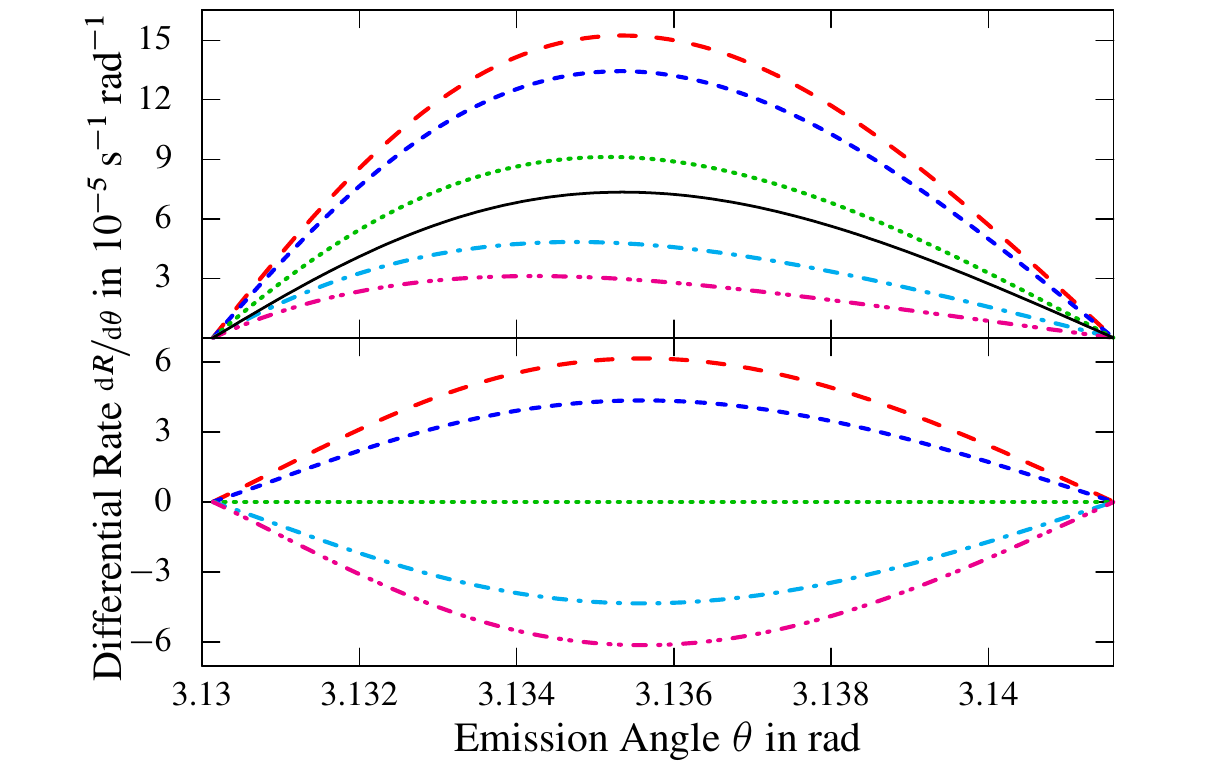}%
}

\subfigure[\figlab{boost-ph-sumint-135}
Total photon energy (in nuclear rest frame) of $1.35 \unit{MeV}$
]{%
\includegraphics[width=\columnwidth]{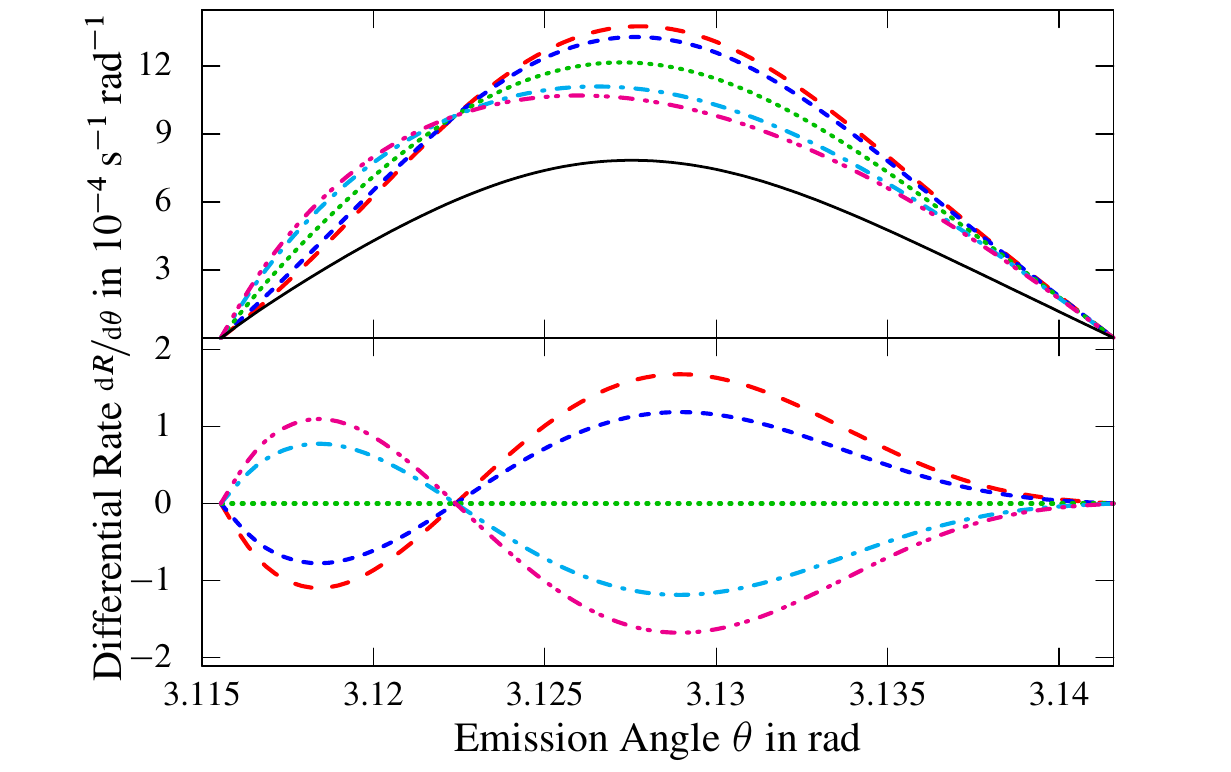}%
}

\caption{\figlab{boost-ph-sumint}(Color online)~
Phase variation in the sums of all terms (top panels) and the interference contributions (bottom panels) of the angular-differential partial rates of laser pair $(2, 4)$ for two energies  transformed to the laboratory frame ($\gamma = 50$). The sum of the direct terms as in \figref{boost-ph-subsum} is given for reference --
Notation, legend, and parameters are identical to those in \figref{ph-sumint}.
}

\end{center}
\end{figure}

Note that we do not give a figure for the variation of the nuclear Lorentz-factor $\gamma$.
The main influence of a different $\gamma$, given that the photon energies for the two modes are adjusted so that the total photon energy stays the same, would be a different emission angle range. The higher the $\gamma$, the more narrowed to the right of the spectrum the kinematically allowed values of $\theta$ would be.
In general, for an appropriately high $\gamma$, which means larger than about $5$, the width of the allowed $\theta$-range is proportional to $\nicefrac{1}{\gamma}$.
The form of the angular distribution, on the other hand, is practically unchanged from a $\gamma$ as small as about $2$ onwards.

\section{Summary and Conclusion} \seclab{concl}

In our study on how interference between two laser modes, with both intensity parameters $\xi_i \ll 1$, affects the rates of electron-positron pair creation, we investigated how the rates change under variation of various parameters.
We began with a variation of the energy of a single photon and thus the number of photons needed to provide the energy to overcome the pair-creation threshold.
Then a variation of the total photon energy for a fixed number of photons was considered, effectively changing the excess energy available as kinetic energy for the created particles.
Finally, we discussed the variation of the relative phase between the two laser modes, i.e., the lateral offset between the peaks of the two laser wave amplitudes.
This was investigated for differential partial rates, discussing individual terms in a four-index sum introduced by Fourier expansion of all periodic functions in the squared pair-creation amplitude, and for differential rates, where the first was given for better illustration and the latter as an actual physical observable. Additionally, the effects on total rates were discussed.

We found only laser pairs with even minimal photon numbers for both modes to show contributions from interference terms. Of those, laser pair $(2, 4)$ yields the strongest interference contribution
%, even though combination $(1,2)$ with lower photon number but identical frequency ratio does not feature any. Thus we used combination $(2, 4)$ 
and was thus used as the prime example for the subsequent studies.
From the variation of the total photon energy for this laser pair, we found that for higher energies the interference contribution decreases, although a change in the shape of the angular distribution leads to a more pronounced peak position shift along the emission angle axis once phase variation is considered.
For the lower energies phase variation mainly manifests in a raised (lowered) summed-up differential rate and thus eventually in an increase (decrease) of the total rate.
The latter effect can be closely related to the phase dependence of the peak electric field.

Finally, the results were transferred to the laboratory frame, where both effects -- peak height and position change -- can be found as well, even though the peak emission angle shift is over a much narrower region, as the whole angular spectrum is Lorentz contracted.

Considering a potential experimental realization, the necessary photon energies in the X-ray regime are accessible today with free-electron lasers (FELs), such as the Linac Coherent Light Source (LCLS) at SLAC, where, recently, a study on the production of bichromatic laser fields has been performed \cite{lutman} and new techniques such as self-seeding allow strongly increased intensities for a given photon energy \cite{amann}.
In combination with a powerful ion accelerator, these developments may lead to an experimental test of our results.

\begin{acknowledgments}
Funding for this project by the German Research Foundation (DFG) under Grant No. MU 3149/1-1 is gratefully acknowledged.
S.\,A. also wishes to thank the Heidelberg Graduate School of Fundamental Physics (HGSFP) for the generous travel support.
\end{acknowledgments}

%\nocite{*}
\bibliography{b1}

\end{document}